\documentclass[]{aastex62}

\newcommand{\MS}{\ifmmode{\,}\else\thinspace\fi{\rm M}\ifmmode_{\odot}\else$_{\odot}$\fi}
\newcommand{\LS}{\ifmmode{\,}\else\thinspace\fi{\rm L}\ifmmode_{\odot}\else$_{\odot}$\fi}
\newcommand{\RS}{\ifmmode{\,}\else\thinspace\fi{\rm R}\ifmmode_{\odot}\else$_{\odot}$\fi}

\shorttitle{Mass and evolutionary status of a type-II Cepheid}
\shortauthors{Pilecki et al.}

\begin{document}


\title{The dynamical mass and evolutionary status of the type-II Cepheid in the eclipsing binary system OGLE-LMC-T2CEP-211 with a double-ring disk
\footnote{Based on observations collected at the European Southern Observatory, Chile}
\footnote{This paper includes data gathered with the 6.5m Magellan Clay Telescope at Las Campanas Observatory, Chile.}
}

\correspondingauthor{Bogumi{\l} Pilecki}
\email{pilecki@camk.edu.pl}

\author[0000-0003-3861-8124]{Bogumi{\l} Pilecki}
\affiliation{Centrum Astronomiczne im. Miko{\l}aja Kopernika, PAN, Bartycka 18, 00-716 Warsaw, Poland}

\author{Ahmet Dervi{\c s}o{\u g}lu}
\affiliation{Department of Physics, University of Zagreb, Bijeni{\v c}ka cesta 32, 10000 Zagreb, Croatia}
\affiliation{Department of Astronomy \& Space Sciences, Erciyes University, Kayseri, Turkey}

\author{Wolfgang Gieren}
\affiliation{Universidad de Concepci{\'o}n, Departamento de Astronom{\'i}a,Casilla 160-C, Concepci{\'o}n, Chile}
\affiliation{Millenium Institute of Astrophysics, Santiago, Chile}

\author{Rados{\l}aw Smolec}
\affiliation{Centrum Astronomiczne im. Miko{\l}aja Kopernika, PAN, Bartycka 18, 00-716 Warsaw, Poland}

\author{Igor Soszy{\'n}ski}
\affiliation{Obserwatorium Astronomiczne Uniwersytetu Warszawskiego, Al. Ujazdowskie 4, 00-478 Warsaw, Poland}

\author{Grzegorz Pietrzy{\'n}ski}
\affiliation{Centrum Astronomiczne im. Miko{\l}aja Kopernika, PAN, Bartycka 18, 00-716 Warsaw, Poland}

\author{Ian B. Thompson}
\affiliation{Carnegie Observatories, 813 Santa Barbara Street, Pasadena, CA 91101-1292, USA}

\author{M{\'o}nica Taormina}
\affiliation{Centrum Astronomiczne im. Miko{\l}aja Kopernika, PAN, Bartycka 18, 00-716 Warsaw, Poland}

\begin{abstract}

We present the analysis of a peculiar W~Virginis (pWVir) type-II Cepheid, OGLE-LMC-T2CEP-211 ($P_{puls}=9.393\,d$), in a double-lined binary system ($P_{orb}=242\,d$), which shed light on virtually unknown evolutionary status and structure of pWVir stars. The dynamical mass of the Cepheid (first ever for a type-II Cepheid) is $0.64\pm{}0.02\,M_\odot$ and the radius $R=25.1\pm{}0.3\,R_\odot$. The companion is a massive ($5.67\,M_\odot$) main-sequence star obscured by a disk. Such configuration suggests a mass transfer in the system history. We found that originally the system ($P_{orb}^{init}=12\,d$) was composed of $3.5$ and $2.8\,M_\odot$ stars, with the current Cepheid being more massive. The system age is now $\sim{}$200 My, and the Cepheid is almost completely stripped of hydrogen, with helium mass of $\sim{}92\%$ of the total mass. It finished transferring the mass 2.5 My ago and is evolving towards lower temperatures passing through the instability strip. Comparison with observations indicate a reasonable $2.7\cdot{}10^{-8}\,M_\odot/y$ mass loss from the Cepheid. The companion is most probably a Be main-sequence star with $T=22000\,K$ and $R=2.5\,R_\odot$. Our results yield a good agreement with a pulsation theory model for a hydrogen-deficient pulsator, confirming the described evolutionary scenario. We detected a two-ring disk ($R_{disk}\sim\,116\,R_{\odot}$) and a shell ($R_{shell}\sim\,9\,R_{\odot}$) around the companion, that is probably a combination of the matter from the past mass transfer, the mass being lost by the Cepheid due to wind and pulsations, and a decretion disk around a rapidly rotating secondary. Our study together with observational properties of pWVir stars suggests that their majority are products of a similar binary evolution interaction.

\end{abstract}


\section{Introduction}
\label{sec:intro}

Type-II Cepheids are low-mass pulsating stars that belong to the disc and halo populations \citep{2002PASP..114..689W}. They are a much older counterpart of the more massive classical Cepheids with periods and amplitudes in a similar range but about 1.5-2 mag fainter. They exhibit a tight and well-defined period-luminosity (P-L) relation \citep{1908AnHar..60...87L} and may serve as  good distance indicators, allowing a measurement of distances both inside and outside of our Galaxy \citep{2009AcA....59..403M,2011MNRAS.413..223M,2017A&A...604A..29G,2018AcA....68...89S}.

Compared to classical Cepheids, our knowledge of type-II Cepheids is very poor, being more qualitative than quantitative. Type-II Cepheids are usually divided into three subgroups depending on the pulsation period, observational properties and evolutionary status but with a similar period-luminosity relation.
Starting from the work of \citet{1976ApJ...204..116G,1985MmSAI..56..169G} it is generally accepted that those with the shortest periods (called BL Herculis or BL Her stars) are evolving from the Horizontal Branch to the Asymptotic Giant Branch (AGB). Those with periods in the range of 4-20 days (called W Virginis or W Vir stars) are on their way up the AGB and enter the pulsational instability strip due to helium shell flashes, which makes them move to higher temperatures on the HR diagram. Finally, those with the longest periods (called RV Tauri or RV Tau stars) are leaving the AGB on their way to the white dwarf cooling sequence. The result of Gingold's work was only qualitative though, explaining the difference between these three subgroups with different periods, without an explanation of the relative rate of occurrence nor a measurement of the values of basic parameters including the masses of these variables.
The situation is even more complicated as more recent evolutionary models \citep[see][and references therein]{2016CoKon.105..149B} do not predict thermal pulses that would explain the existence of W Vir stars.

It is clearly important to obtain direct measurements of the masses for a sample of type-II Cepheids to pinpoint their evolutionary status. The best means for such measurements are observations of eclipsing binary systems in which one or both components are pulsating stars. We have applied this method to eclipsing binaries containing classical Cepheids obtaining very precise masses (\citealt{cep227mnras2013, cep2532apj2015, allcep_pilecki_apj2018}; \citealt{cep9009apj2015}).

As part of a program studying the  characteristics of type-II Cepheids, the OGLE project \citet{2008AcA....58..293S, 2018AcA....68...89S} identified a group of W Virginis stars that for similar periods have different looking light curves, with the ascending branch being steeper than the descending one. They were called peculiar W Virginis stars (hereafter also pWVir). In general, these stars also lie above the normal P-L sequence for type-II Cepheids and are bluer in color. Since a significant fraction of pWVir stars show eclipses and ellipsoidal modulations, it has been  suggested that all of them are members of binary systems. The extra light from the companion would then move the stars above the sequence. Peculiar WVir also have a different spatial distribution from the other type-II Cepheids in the LMC, which suggests that their evolutionary history is different.

Recently we analyzed photometric and spectroscopic observations of a system with one such Cepheid, OGLE-LMC-T2CEP-098 \citep{t2cep098apj2017}. The results indicated that this star, with a mass of about 1.5 solar masses, is not a member of any of the known Cepheid groups,   and is probably a product of a mass reversal during its evolution. Preliminary results for other systems suggest that high mass ratios and the presence of a disk are very common for type-II Cepheids in binary systems \cite{2017EPJWC.15207007P}.

Here we present the analysis of spectroscopic and photometric data for another such system,  OGLE-LMC-T2CEP-211, which was found in the data of the OGLE-4 stage of the OGLE project \citep{ogle2015udalski,2018AcA....68...89S}. We show evidence that the Cepheid is a genuine member of the observed eclipsing binary system. Section 2 presents the data used in this study, Section 3 presents our analysis and the results, Section 4 contains an analysis of the evolution of this binary system, Section 5 discusses the pulsational properties of the Cepheid. We conclude in Section 6  with a summary of the paper together with a discussion of the results.

\section{Data}
\label{sec:data}

\begin{figure}
\begin{center}
  \resizebox{0.4\linewidth}{!}{\includegraphics{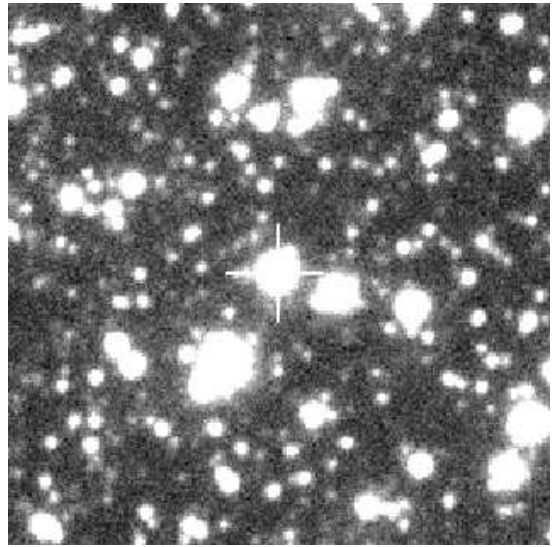}} \\
\caption{Neighborhood of OGLE-LMC-T2CEP-211 (marked with a white cross). An area of 60x60 arcseconds is shown.
\label{fig:8280_stars}}
\end{center}
\end{figure}

Our analysis makes use of 797 measurements in the Cousins I-band collected with the Warsaw telescope at Las Campanas Observatory and provided by the OGLE project \citep{2018AcA....68...89S}. The OGLE V-band data were only used to determine the system brightness and color, as the most important (see Section~\ref{sub:modeling} for explanation) secondary eclipse is not covered in this band.
Basic photometric data for the system are given in Table~\ref{tab:basic}. A finder chart of the system is shown in Fig.~\ref{fig:8280_stars}.

The spectroscopic data were acquired using the MIKE spectrograph on the 6.5-m Magellan Clay telescope at Las Campanas Observatory and the UVES spectrograph on VLT at Paranal Observatory in Chile. In total we obtained 25 high-resolution spectra (19 MIKE + 6 UVES). The MIKE data were reduced using Daniel Kelson's pipeline available at the Carnegie Observatories Software Repository\footnote{http://code.obs.carnegiescience.edu/}. The  UVES data were reduced using ESO Reflex software and the official pipeline available at the ESO Science Software repository\footnote{http://www.eso.org/sci/software.html}.

The photometric data are available from the OGLE webpage, and our radial velocity measurements are available on the webpage\footnote{http://users.camk.edu.pl/pilecki/p/t2c211} and in the online version of the manuscript.

\begin{deluxetable}{lcc}
\tablecaption{Basic data of the OGLE-LMC-T2CEP-211 system.\label{tab:basic}}
\tablewidth{0pt}
\tablehead{
\colhead{Parameter} & \colhead{Value} & \colhead{Unit}
}
\startdata
Right Ascension     & 04:46:27.44    & hh:mm:ss.s \\
Declination         & -70:59:16.7    & dd:mm:ss.s \\
$W_I = I_{C} - 1.55(V-I_{C})$ & 14.747       & mag        \\ 
$I_{C}$             & 16.088         & mag        \\ 
$V$                 & 16.953         & mag        \\ 
$(V-I_{C})$         &  0.865         & mag        \\ 
$P_{orb}$           & 242.18(10)     & days       \\ 
$P_{puls}$          & 9.3933(5)      & days       \\
\enddata
\tablecomments{Based on the OGLE project data \citep{2018AcA....68...89S}. Periods are calculated from the separated light curves. Magnitudes are flux-averaged over pulsation cycle.}
\end{deluxetable}

\pagebreak

\section{Analysis and results} \label{sec:analysis}

\subsection{Photometry}
\label{sub:photo}

The brightness and light curve of the pulsating star suggests that this object is a type II Cepheid. It is classified as a peculiar W Virginis star in the OGLE collection of variable stars \citep{2018AcA....68...89S}. Here  we present an analysis of the light curve, including a separate confirmation of the variability type of the star.

\begin{figure}
\begin{center}
  \resizebox{0.6\linewidth}{!}{\includegraphics{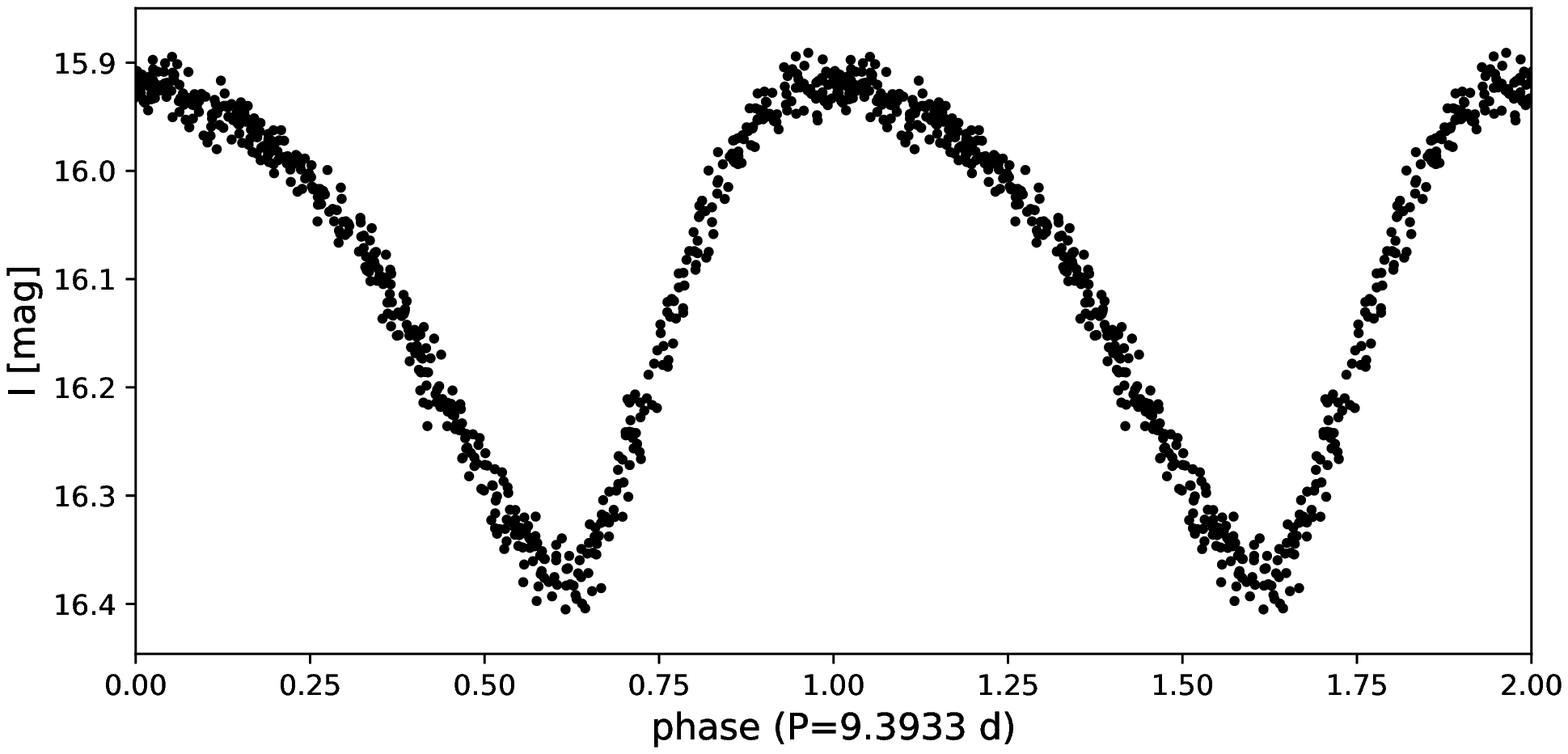}} \\
  \resizebox{0.6\linewidth}{!}{\includegraphics{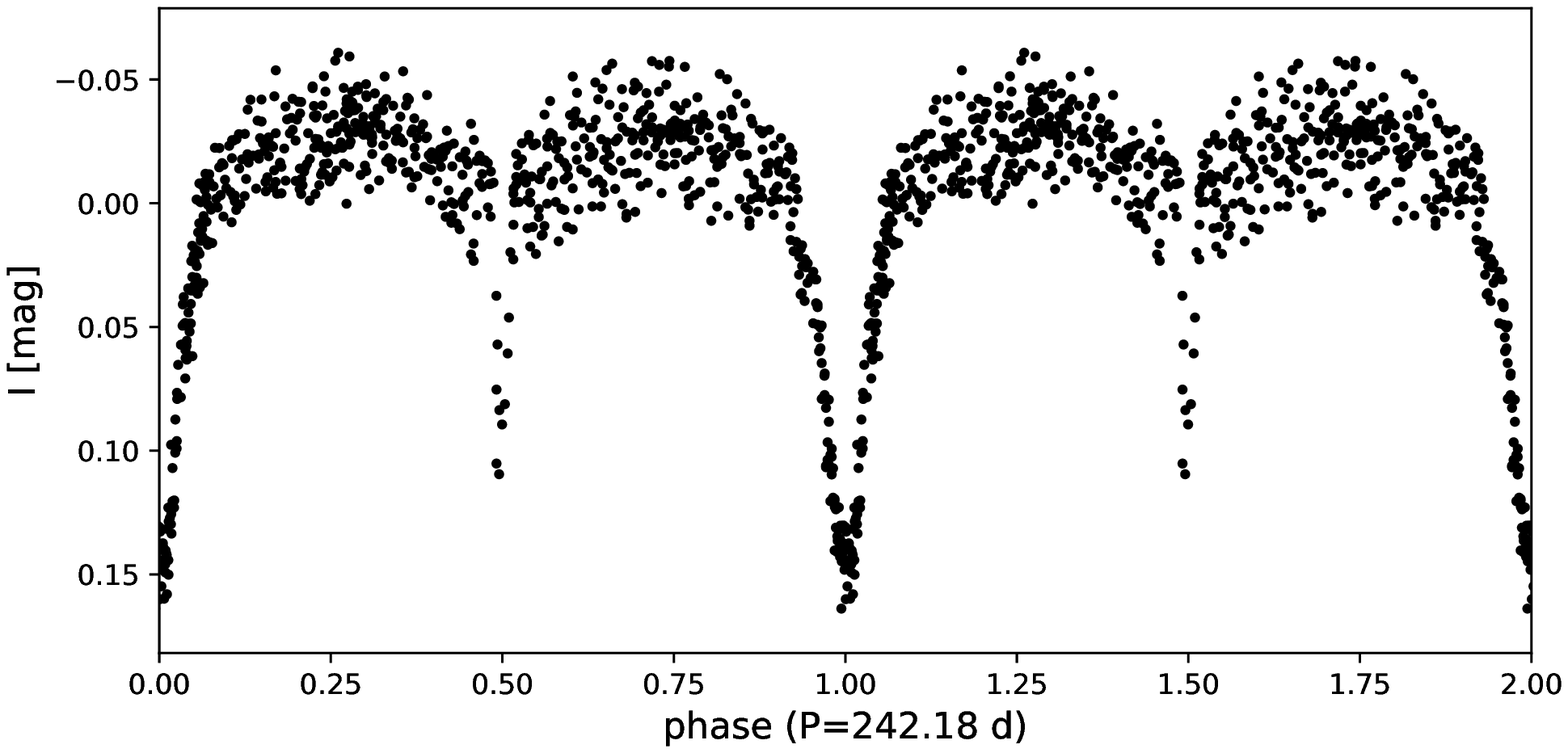}} \\
\caption{{\em top:} Pulsational I-band light curve of the Cepheid OGLE-LMC-T2CEP-211 folded with the ephemeris $T_{max} ($HJD$) = 2456123.1120 + 9.3933\times $E.
$\vert$ {\em bottom:} The I-band eclipsing light curve of the system freed from the pulsation of the Cepheid, folded with the ephemeris $T_{pri} ($HJD$) = 2455997.92  + 242.18\times $E. The primary eclipse is much wider than the secondary eclipse.
\label{fig:8280_lc}}
\end{center}
\end{figure}

We first removed the observations in and close to the eclipses from the original light curve ($LC_{org}$), and then subtracted a low-order Fourier series fitted to the ellipsoidal variability that was still visible in the data (out of eclipses). The light curve was also detrended in the process, i.e. we have subtracted a systematic brightness variation with a period of about one year.
The resulting pulsational light curve ($LC_{puls}$) is presented in the left panel of Fig.~\ref{fig:8280_lc}.
This pulsational variability was then subtracted from the original light curve to obtain the eclipse light curve ($LC_{ecl}$) which is shown in the right panel. The orbital and pulsational periods determined in this process are given in Table~\ref{tab:basic}.

\begin{figure}
\begin{center}
  \resizebox{0.6\linewidth}{!}{\includegraphics{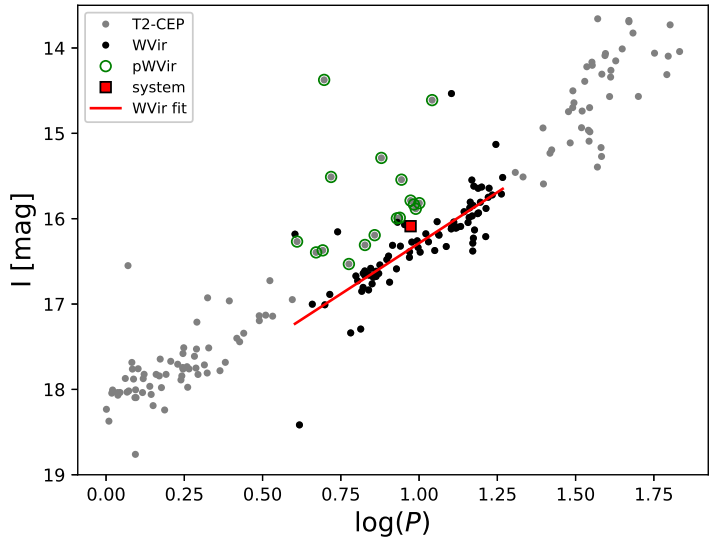}} \\
\caption{Position of the Cepheid in the system studied in this paper, on the period-luminosity diagram for type II Cepheids in the LMC from the OGLE catalog. P-L relation for W Virginis stars is shown. OGLE-LMC-T2CEP-211 is one of the faintest pWVir stars for its period.
\label{fig:8280_perlum}}
\end{center}
\end{figure}

With the observed brightness and period the star falls directly on the period-luminosity relation for type II Cepheids in the part occupied by the W Virginis variables (W Vir) in the LMC -- see Fig.~\ref{fig:8280_perlum}. As seen in Fig.~\ref{fig:8280_lc} the shape of its pulsation light curve is however different, with the ascending branch steeper than the descending one, quite opposite to typical W Vir stars with similar periods.
In the OGLE Collection of Variable stars with this type of light curves\footnote{see also: \texttt{http://ogle.astrouw.edu.pl/atlas/W\_Vir.html}} are classified as peculiar W Virginis stars \citep{2008AcA....58..293S}.

We have studied the shape of the Cepheid $LC_{puls}$ through its Fourier decomposition parameters \citep{1981ApJ...248..291S}. To describe the light curve shape the amplitude ratios $R_{i1} = a_i/a_1$ and phase differences $\phi_{i1} = \phi_i - i*\phi_1$ are used, with a series of the following form fitted to the data:

$$ a_0 + \sum\limits_{i=1}^{n} a_i \cos\left[i\omega(t-t_0) + \phi_i\right] $$

\begin{figure}
\begin{center}
  \resizebox{0.6\linewidth}{!}{\includegraphics{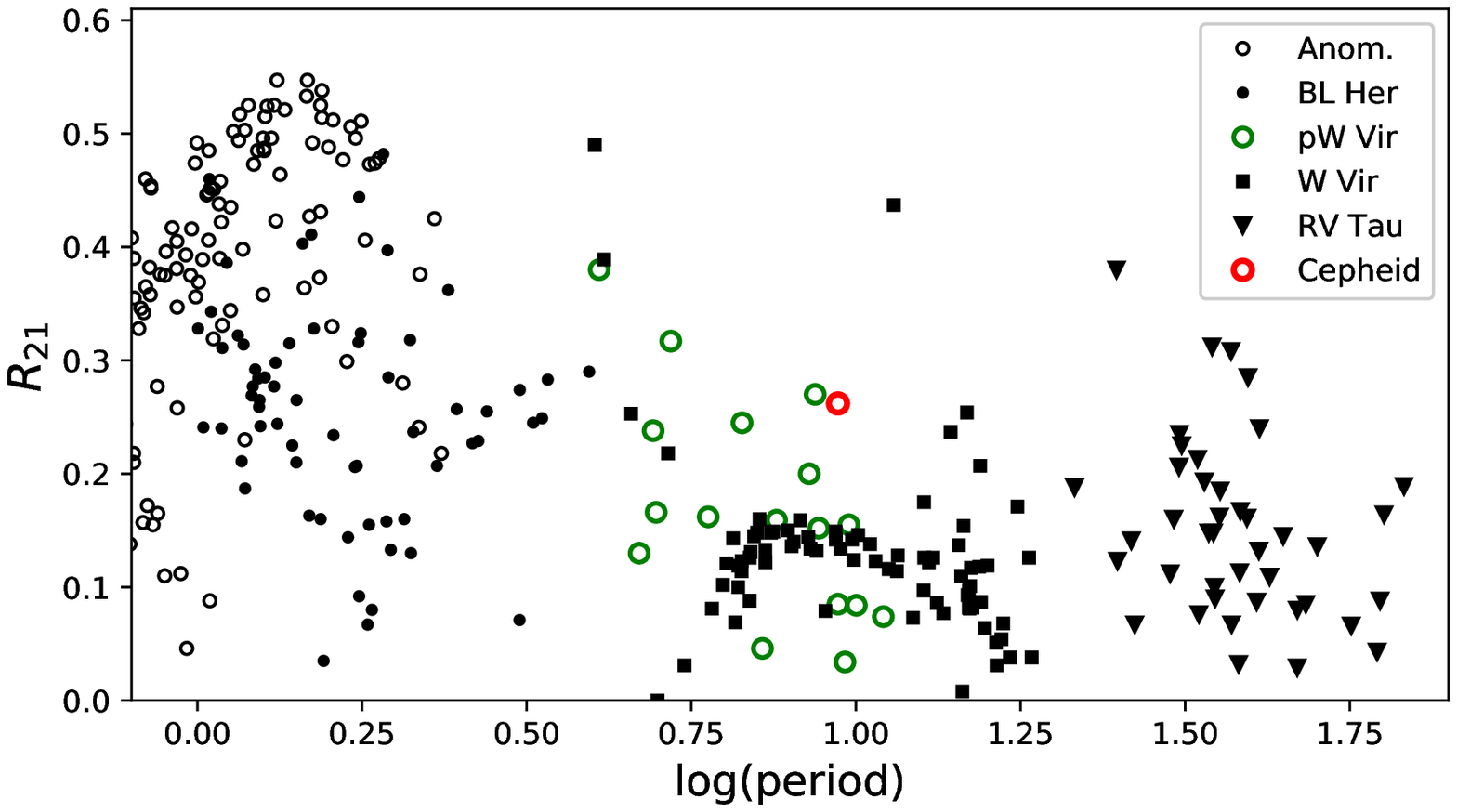}}  \\    
  \resizebox{0.6\linewidth}{!}{\includegraphics{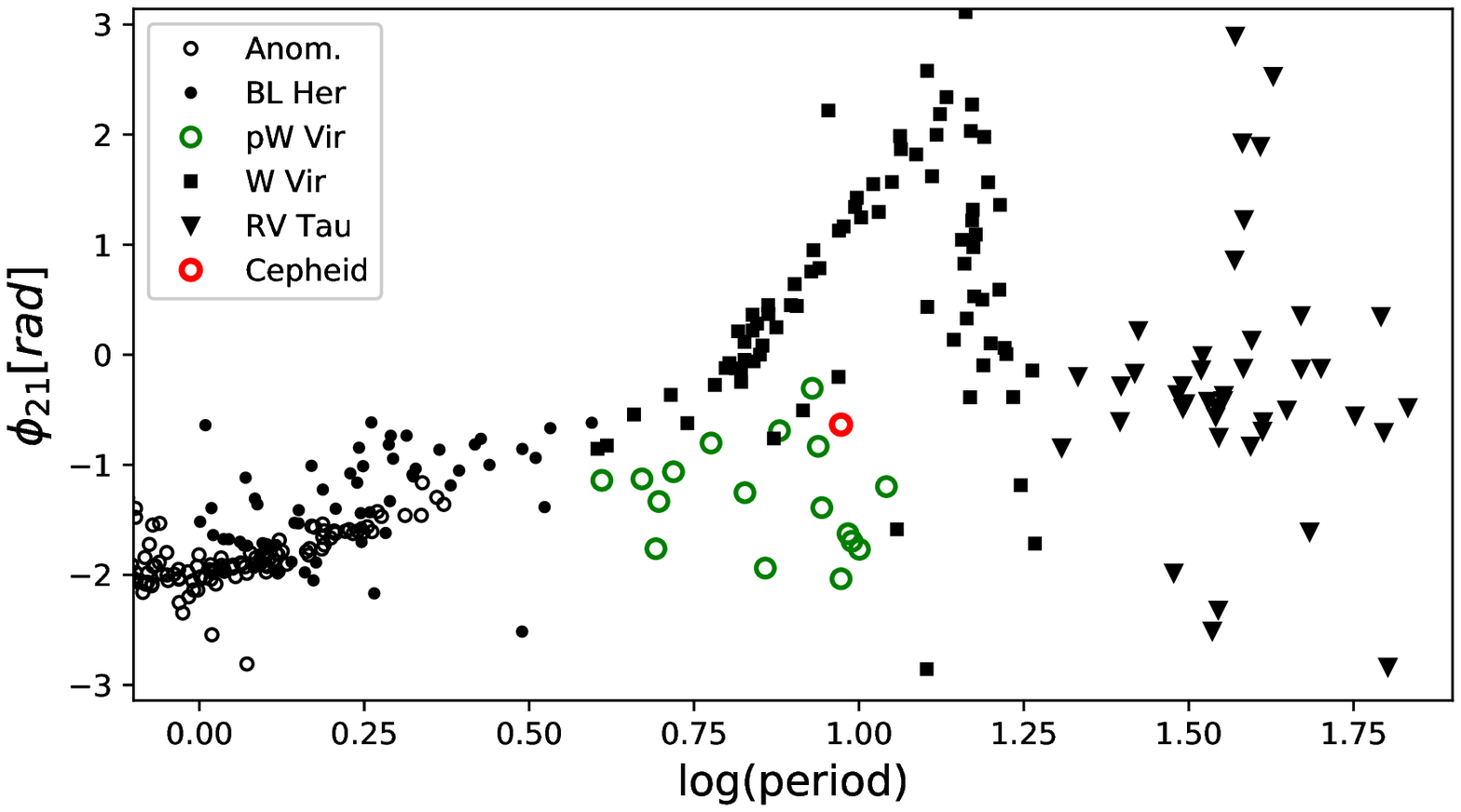}} \\
\caption{Fourier decomposition parameters -- $R_{21}$, $\phi_{21}$ for Type II Cepheids from the OGLE catalog. The position of OGLE-LMC-T2CEP-211 (marked by a red circle) is consistent with the star being of a peculiar W Virginis type. 
\label{fig:8280_fourier21}}
\end{center}
\end{figure}

The position of the star in the $R_{21}$ and $\phi_{21}$ vs. period plane is compared to other Type II Cepheids in Fig.~\ref{fig:8280_fourier21}. It is consistent with the classification as a peculiar W Virginis star.

After the subtraction of both (pulsational and eclipsing) variabilities we have checked if there is any other residual periodic variability. We found two significant peaks in the frequency domain ($0.0982d^{-1}$ and $0.1148d^{-1}$) which are equally spaced from the main pulsation frequency of $0.1065 d^{-1}$.
These additional peaks are shifted by 2/$P_{orb}$ from the main peak, which means that the pulsations are affected by something related to the binarity of the Cepheid. \citet{2010AcA....60...91S} suggested that tidal interactions are involved -- the Cepheid is not spherical and we see the star differently depending on the orbital phase.

We have also detected a significant but irregular pulsation period variation composed of 3-4 erratic short scale episodes of period decreases and increases during the 8 years of observations. The measured average period change is $(2.4 \pm 0.3) \times 10^{-6}$ but this measurement suffers from the irregular variability of the period and highly not uniform distribution of observations -- the cadence is significantly lower after HJD 2456400.  It will be important to re-do this analysis once observations spanning a longer duration are available in order to check the influence of the evolution on the period of the Cepheid. The origin of the short-scale variations is not yet clear.

\subsection{Spectroscopy}
\label{sub:radvel}

\begin{figure}
\begin{center}
  \resizebox{0.5\linewidth}{!}{\includegraphics{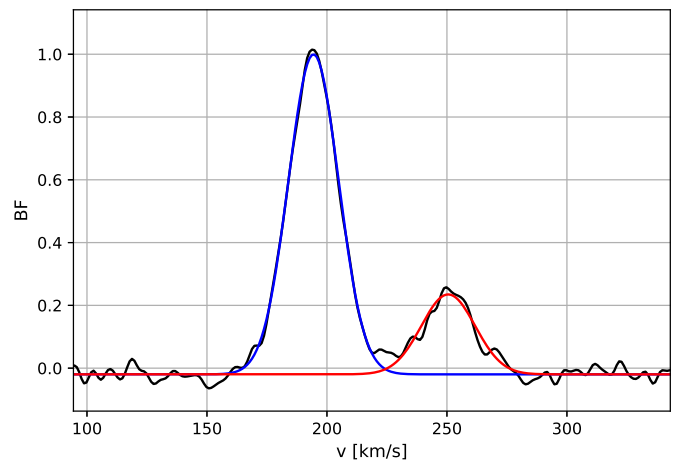}} \\
\caption{Broadening Function profiles for the components of OGLE-LMC-T2CEP-211. The signal from the secondary is weak, but strong enough for a reliable velocity measurement.
\label{fig:8280_sb2}}
\end{center}
\end{figure}

Unlike the OGLE-LMC-T2CEP-098 system with a type II Cepheid we studied before \citep{t2cep098apj2017}, OGLE-LMC-T2CEP-211 is a clear double-lined spectroscopic binary (SB2). The Broadening Function profile of both components is shown in Fig.~\ref{fig:8280_sb2}. Although the signal from the secondary is relatively weak, it is strong enough to be detectable in the majority of our spectra and to derive the velocity of the secondary with good precision. This makes possible a reliable dynamical mass determination.

To measure velocities we used the Broadening Function method \citep{1992AJ....104.1968R, 1999ASPC..185...82R} implemented in the {\tt RaveSpan} application \citep{cep227mnras2013, cep2532apj2015} with templates matching the Cepheid and its companion in the temperature-gravity plane. The templates were taken from the library of theoretical spectra of \citet{2005A&A...443..735C}. Radial velocities were measured in the range 4125 to 6800 $\mathring{A}$. The typical formal errors of the velocities of the system components are about 600 m/s.

The measured velocities show cyclical variations with the pulsational and orbital periods as derived from the light curve analysis. This unambiguously confirms that the Cepheid OGLE-LMC-T2CEP-211 is a component of an eclipsing binary system.

To disentangle the orbital and pulsational motions we simultaneously fitted the eccentricity $e$, argument of periastron $\omega$, velocity semi-amplitudes $K_1$ and $K_2$, systemic velocity $\gamma$, the orbital period $P_{orb}$, the reference time $T_0$, and a number of Fourier series coefficients representing the pulsational radial velocity (RV) curve together with a constant offset from the $\gamma$-velocity (the K-term). The shape of the pulsational RV curve is far from sinusoidal and we had to use 5th order series to describe it. Because of the relatively low number of data points, we had to use regularization \citep{1998_Regularization_Neumaier} to avoid over-fitting.

The time span of spectroscopic and photometric observations is similar and the orbital cycles more uniformly covered in the radial velocity curve than in the light curve so we decided to fit the orbital period independently, but the same value of $P_{orb}$ was derived from both data sets within the uncertainties.
Because of the much better phase coverage of the pulsational light curve, we decided to use the period of the pulsation taken from the photometric data analysis, but fit the reference time of the maximum to account for any phase shift. If we however leave the period as a free parameter the resulting value is perfectly consistent with the mean period from the light curve. As the spectroscopic observations were obtained in more recent epochs, we conclude that on average the period change is insignificant on the time scale of the observations, although short-scale variations may be quite rapid.

Small eccentricity ($e=0.010 \pm 0.015$) was initially obtained for the best model. However, as the value is consistent with zero, for simplicity we eventually assumed a circular orbit for the system.

\begin{figure}
\begin{center}
  \resizebox{0.6\linewidth}{!}{\includegraphics{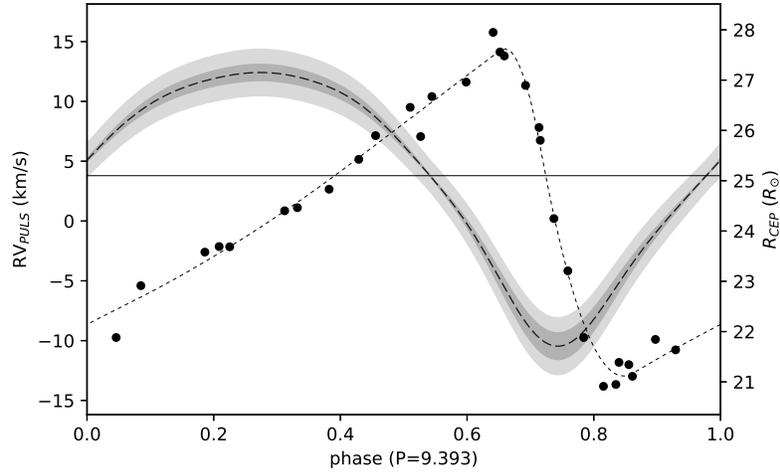}} \\
\caption{Pulsational radial velocity curve (points) with a spline fit (dotted line) and the radius variation (dashed line) of the Cepheid over one pulsation cycle. The dark gray area marks the p-factor range of 1.2-1.4, while the light gray area represents the  uncertainty in the radius determination. The amplitude of the radius variation is $5.0-5.8 R_\odot$ depending on the assumed p-factor.
\label{fig:8280_rvpuls}}
\end{center}
\end{figure}

\begin{figure}
\begin{center}
  \resizebox{0.6\linewidth}{!}{\includegraphics{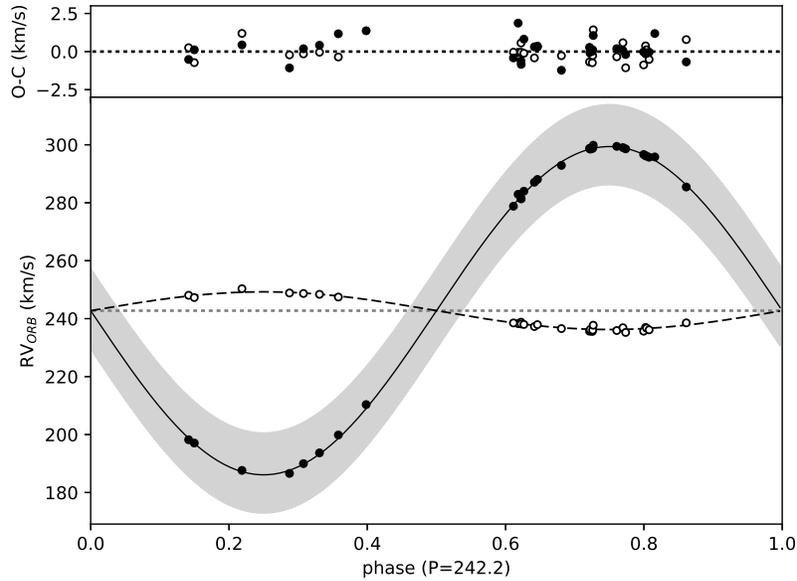}} \\
\caption{Measured orbital radial velocities of the Cepheid (points) with the pulsation velocity removed over-plotted on the orbital solution. The pulsational variability range of the Cepheid is marked with the light gray area.
\label{fig:8280_rvorb}}
\end{center}
\end{figure}

The final model parameters are presented in Table~\ref{tab:8280_spec} and the pulsational and orbital RV curves are shown in Fig.~\ref{fig:8280_rvpuls} and \ref{fig:8280_rvorb}, respectively.  The pulsational amplitude is about 27.5 km/s, which for the pulsational period of 9.4 days translates to an amplitude of the radius change of 5 to 5.8 $R_\odot$ depending on the assumed p-factor ranging over   1.2--1.4 \citep{Kervella2017_RS_Pup,2017A&A...608A..18G}.

\begin{deluxetable}{lr@{ $\pm$ }lc}
\tablecaption{Orbital solution for OGLE-LMC-T2CEP-211. \label{tab:8280_spec}}
\tablewidth{0pt}
\tablehead{
\colhead{Parameter} & \multicolumn{2}{c}{Value}  & \colhead{Unit}
}
\startdata
\hspace{0.2cm}$T_0$           & 2455755.5     &    0.3   &  \hspace{0.2cm}days\hspace{0.2cm}  \\
\hspace{0.2cm}$P_{orb}$       &   242.24      &   0.08   &  days    \\
\hspace{0.2cm}$\gamma$        &   242.31      &   0.05   &  km/s    \\    
\hspace{0.2cm}k-term$^a$      &   -0.8        &   0.2    &  km/s    \\ 
\hspace{0.2cm}$K_1$           &   56.65       &   0.2    &  km/s    \\
\hspace{0.2cm}$K_2$           &    6.41       &   0.10   &  km/s    \\
\hspace{0.2cm}$e$             &    0.0$^b$    &   0.015  &  -       \\
\hspace{0.2cm}$a \sin i$      &  302.0        &   1.1    &  $R_\odot$ \\ 
\hspace{0.2cm}$q = m_2/m_1$   &    8.82       &   0.14   &          \\
\hspace{0.2cm}$m_1 \sin^3 i$  &    0.640      &   0.013  &  $M_\odot$ \\
\hspace{0.2cm}$m_2 \sin^3 i$  &    5.65       &   0.06   &  $M_\odot$ \\
\hspace{0.2cm}$P_{puls}$      &    9.393      &   0.001  &  days     \\
\enddata
\tablenotetext{a}{The apparent shift of the Cepheid average velocity in reference to the systemic velocity $\gamma$.}
\tablenotetext{b}{For the best model we obtained $e=0.010 \pm 0.015$). For simplicity we fixed the eccentricity at zero.}
\end{deluxetable}

\begin{figure}
\begin{center}
  \resizebox{0.49\linewidth}{!}{\includegraphics{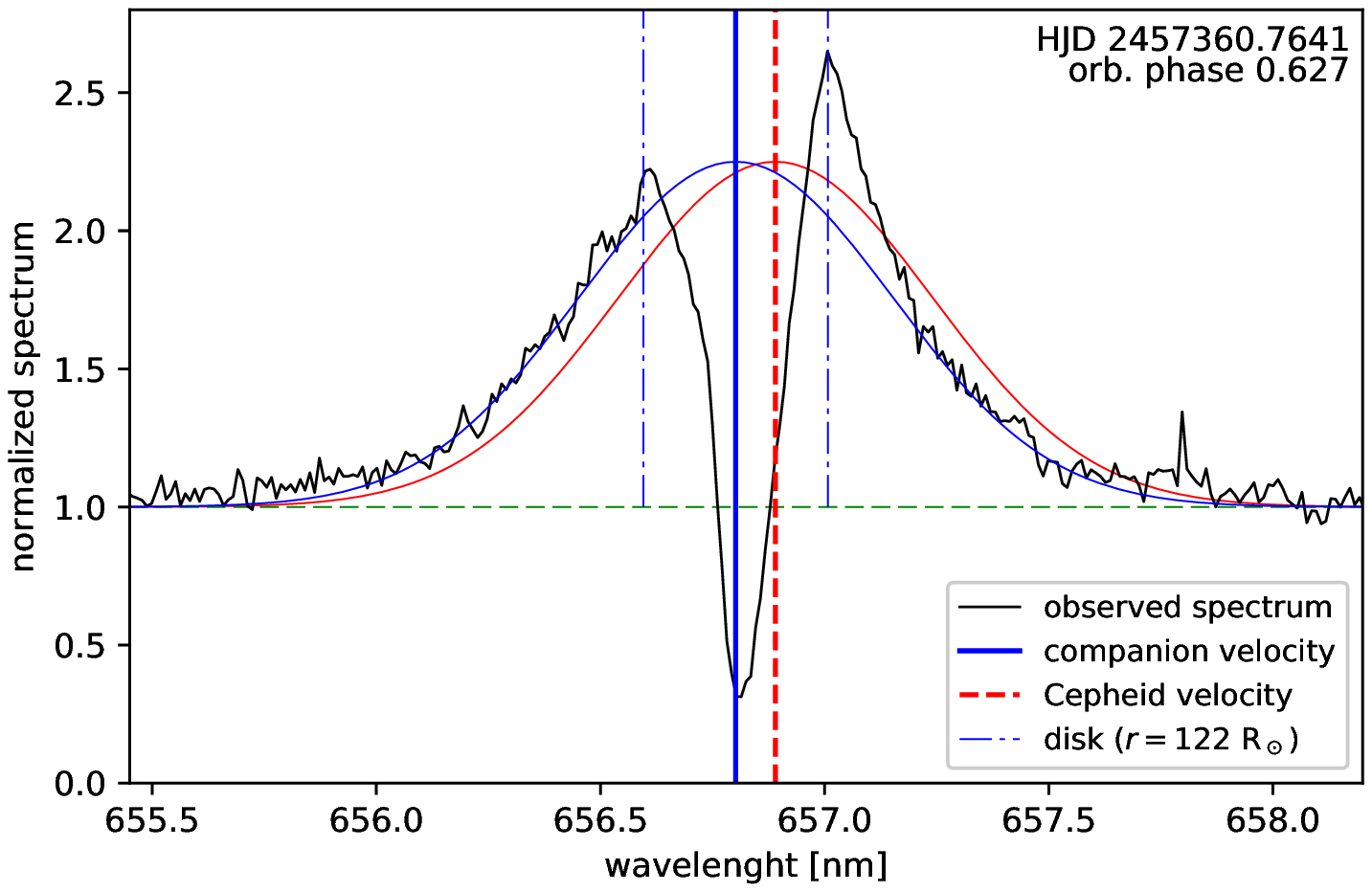}}
  \resizebox{0.49\linewidth}{!}{\includegraphics{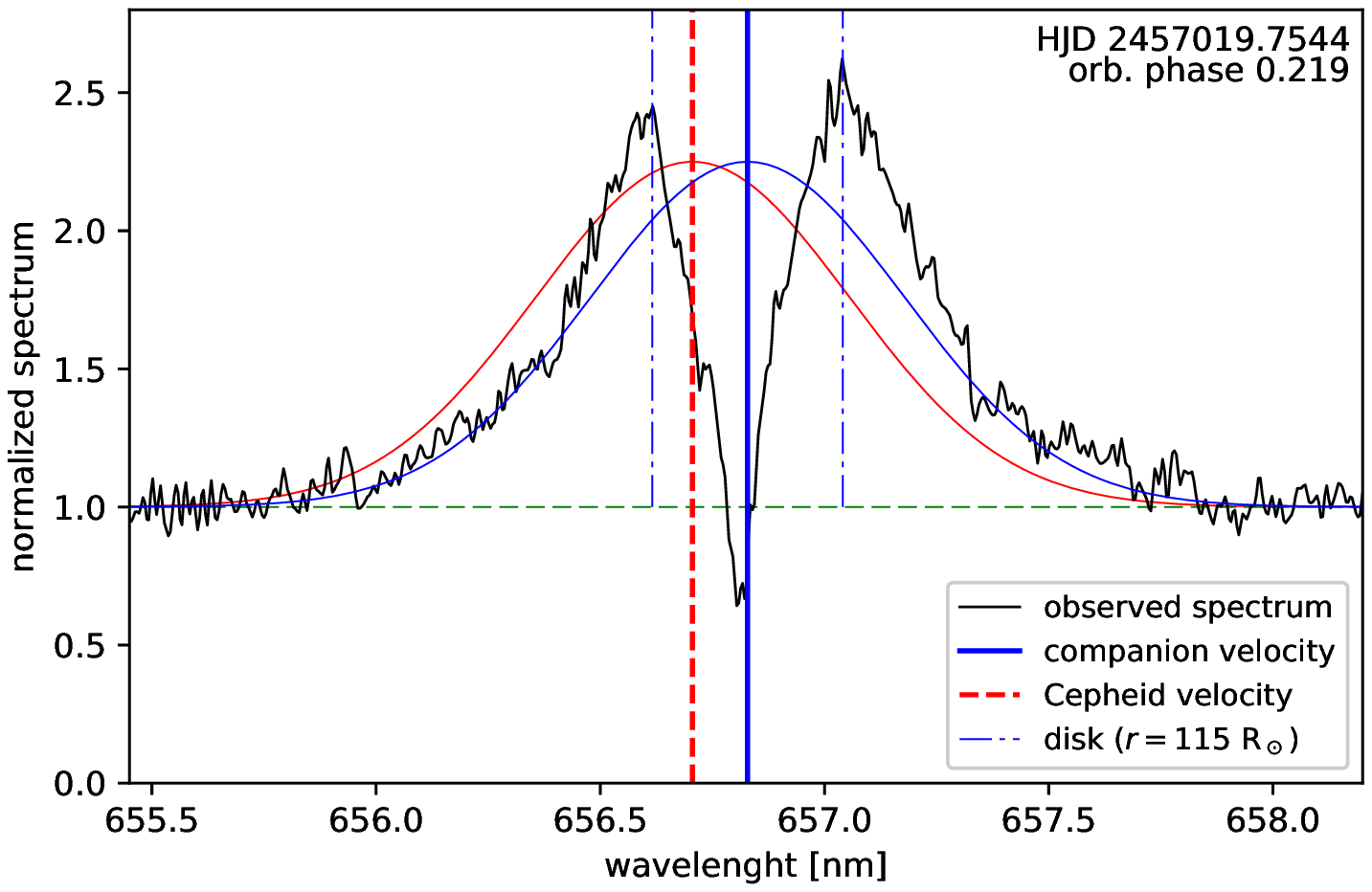}}  
\caption{H$_\alpha$ region for spectra taken at two different orbital phases. The splitting is clearly seen.  Normal profiles ($\sigma$=160 km/s) around components velocities are over-plotted to make it easier to identify the line sources. The broad emission line is clearly correlated with the companion and comes from the circumstellar disk. Disk velocities at $r=122$ $R_\odot$ in the left panel and at $r=115$ $R_\odot$ in the right panel (corresponding with the emission maxima) are marked with dash-dotted lines.
\label{fig:8280_halpha}}
\end{center}
\end{figure}

\begin{figure}
\begin{center}
  \resizebox{0.49\linewidth}{!}{\includegraphics{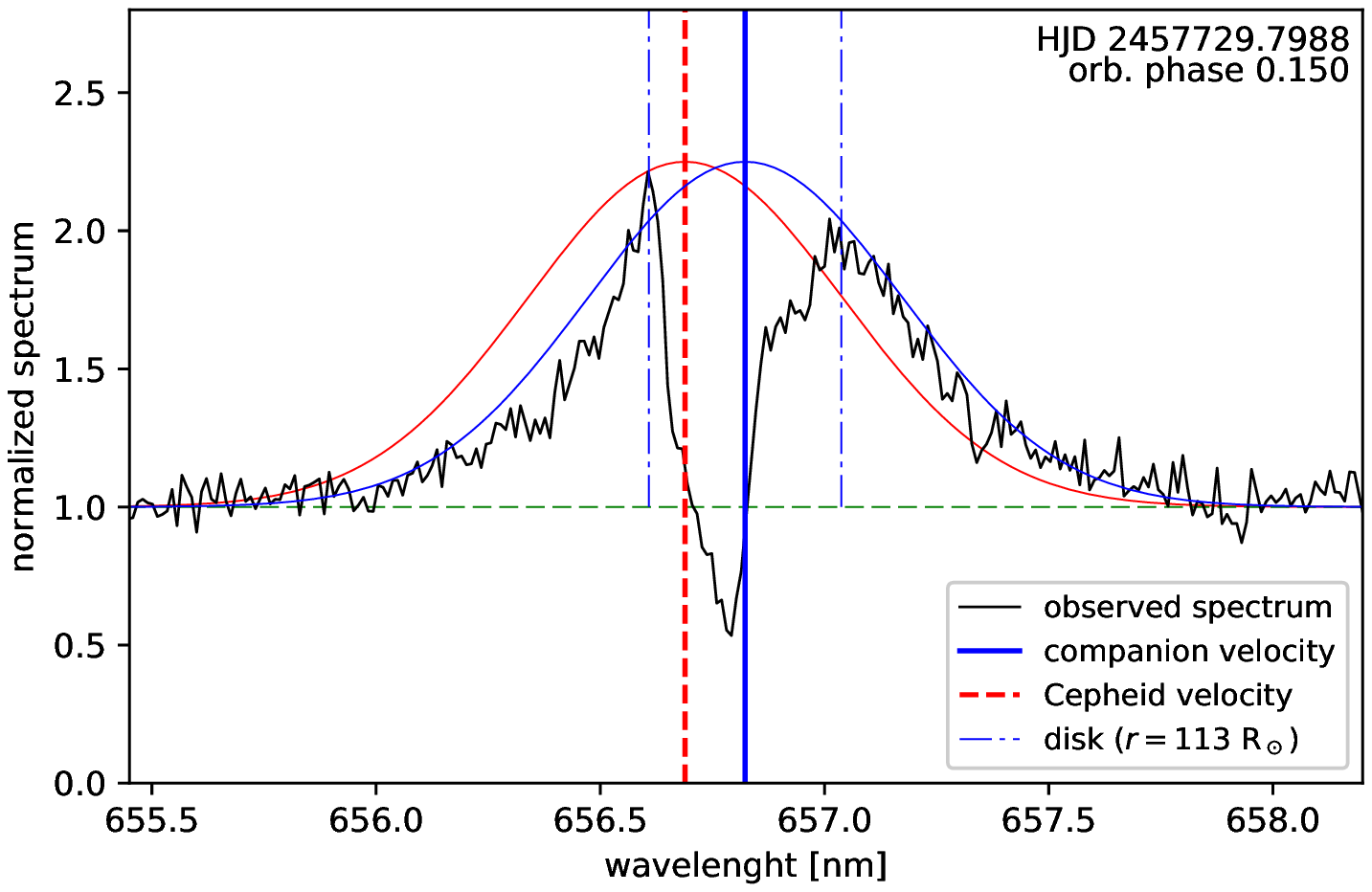}}
  \resizebox{0.49\linewidth}{!}{\includegraphics{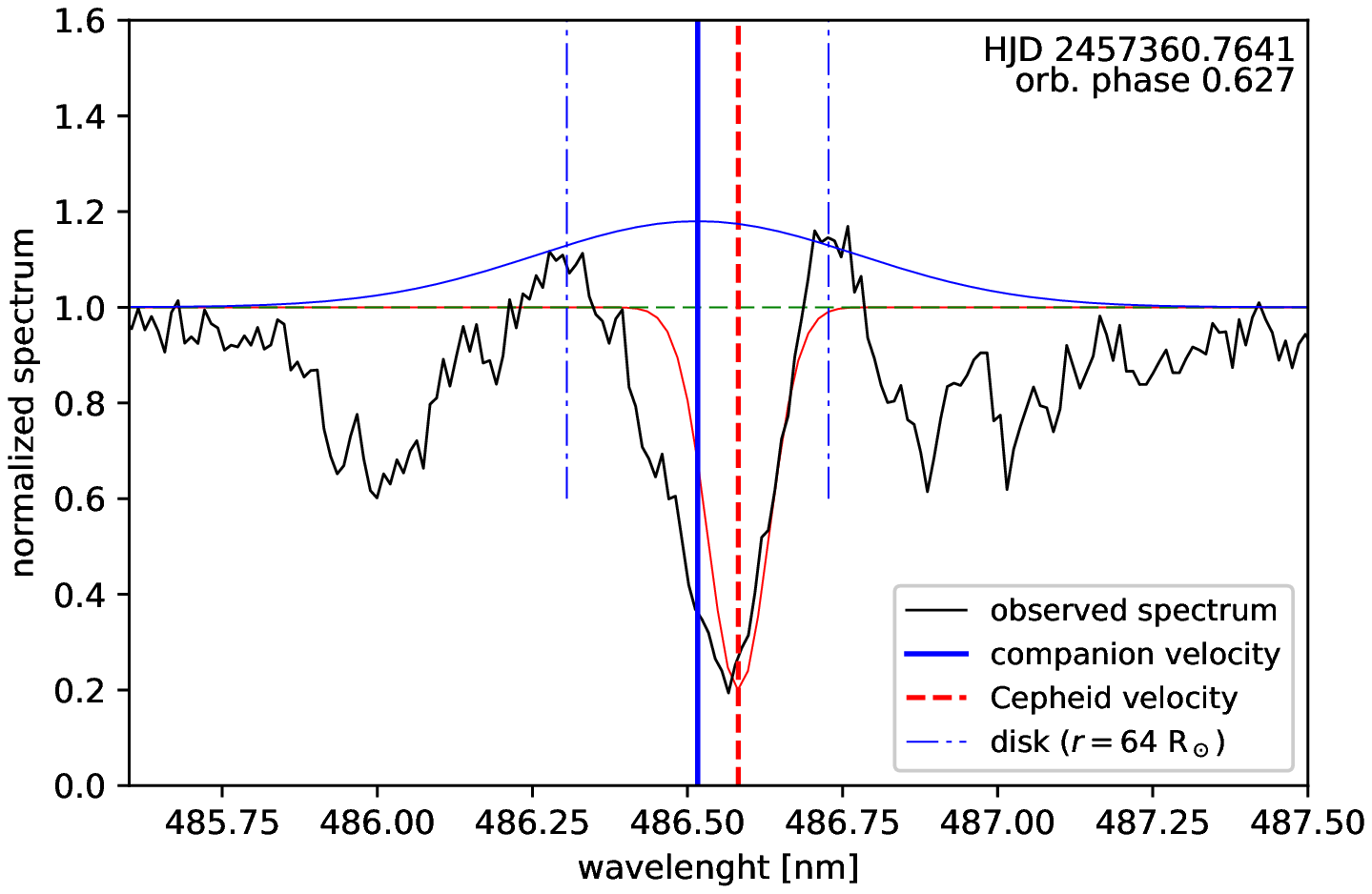}}
\caption{{\em left:} Same as Fig.~\ref{fig:8280_halpha} for a different orbital phase. $\vert$ {\em right:} Similar plot but for H$_\beta$ region. The Gaussian profiles are arbitrarily adjusted. The H$_\beta$ emission maxima come from a region of the disk at $r=64R_\odot$, corresponding to a velocity of 130 km/s.
\label{fig:8280_hbeta}}
\end{center}
\end{figure}

In Fig.~\ref{fig:8280_halpha} and \ref{fig:8280_hbeta} we show the H$_\alpha$ region for spectra taken at three different orbital phases. Gaussian profiles with FWHM = 375 km/s are overplotted, centered on the measured velocities of the two components at the given phases. The strong emission line is clearly correlated with the companion and not the Cepheid. The line splitting is a clear indication of the presence of an edge-on disk \citep{2013A&ARv..21...69R, 2018ApJ...854..109Z}. This type of profile is also typical for Be shell stars \citep{2014ApJ...795...82S}, suggesting  that the massive companion may be of this stellar type. The central absorption line is blended with the shallow absorption line of the Cepheid and thus slightly shifted from the companion velocity. The broad emission line shows signs of variability, which is also typical for Be shell stars. Assuming the emission indeed comes from the circumstellar disk, we can estimate at what distance it originates using a wavelength shift in reference to the companion. We found that the maxima of the emission are shifted by 94-98 km/s, which correspond to a distance of 113 $R_\odot$ to 122 $R_\odot$ from the companion center.

The H$_\beta$ line is much weaker than H$_\alpha$ and thus more noisy, but its shape is also characteristic for stars with a circumstellar disk. As seen in the right panel of Fig.~\ref{fig:8280_hbeta} the absorption feature comes from the environment of the Cepheid, but the emission maxima are located symmetrically around the companion velocity shifted by $\pm$130 km/s. This velocity corresponds to the distance of about 64 $R_\odot$ from the companion center. A presence of very wide calcium lines is also easily noticeable in the data. More advanced analysis of the spectra is for now prohibited by relatively low signal to noise ratio for them, but is planned for the future, once better quality data are collected.

\subsection{Modeling} \label{sub:modeling}

We started the analysis of the system by fitting a simple eclipsing binary system model to the $LC_{ecl}$ light curve, i.e. the light curve with the pulsations subtracted. This fit resulted in a very high eccentricity, needed to explain the difference in the eclipse widths, which is totally excluded from the orbital solution (eccentricities higher than 0.04 are highly improbable).
For a circular orbit we could obtain a good fit to the secondary eclipse, but the primary eclipse was much narrower than that observed (the zero eccentricity impose the eclipses to be equal in width).

To explain the observed features, the presence of some additional dark component is necessary in the system which blocks the Cepheid light around the phase of its eclipse, but having negligible effect on the light curve when the companion is eclipsed. We concluded than a presence of a disk around the  companion to the Cepheid is necessary to explain these observations.

To obtain an approximate solution for  the Cepheid we  cut out the part of the light curve affected by the disc presence (the wider and deeper primary eclipse). Normally in such a case it is not possible to obtain the ratio of the radii. However, as described in \citet{cep2532apj2015} and \citet{t2cep098apj2017}, the presence of a pulsating component in the system adds extra constraints to the model. A light curve solution for such a system together with the orbital solution make it possible to estimate the necessary physical parameters with confidence.

For a detailed description of the method and the analysis steps we refer the reader to \citet{t2cep098apj2017} and \citet{cep227mnras2013}. Here we  present a short summary of the subject.
    
The photometric data were analyzed using a  eclipse modeling tool based on the well-tested {\tt JKTEBOP} code \citep{jktebop2004southworth} modified to allow the inclusion of pulsational variability. The {\tt JKTEBOP} code itself is based on the EBOP code \citep{1981AJ.....86..102P}, in which the stars are treated as spheres for calculating eclipse shapes and as bi-axial ellipsoids for calculating proximity effects. This treatment of the stars makes it useful only for well separated components, but the code is very efficient and has low numerical noise. 

In our approach we generate a two-dimensional light curve that consists of purely eclipsing light-curves for different pulsating phases. A one-dimensional light curve is then generated (interpolated from the grid) using a combination of pulsational and orbital phases calculated with the Cepheid and system ephemerides.

From the photometric solution we have the period, the time of the primary minimum ($T_{I}$), the inclination ($i$), the fractional radii ($r_1$ and $r_2$) and the surface brightness ratios ($j_{21}$). Fractional radii are expressed in the units of $A$, the separation of the components, such that $r_i = R_i/A$, where $R_i$ are the physical radii of the components. The eccentricity ($e$) and the argument of periastron ($\omega$) were taken from the orbital solution. The third light was assumed to be negligible and the p-factor was set to 1.25, an appropriate approximation given the quality of the data.

As  mentioned, the analyzed data set makes it impossible to reliably derive both the primary and secondary radii. Using pulsation theory models as described in \citet{t2cep098apj2017} and the calculated mass and estimated temperature (described below), we  obtain a Cepheid radius for which the calculated and observed periods match. This yields a radius value for the Cepheid of $25.1 \pm 0.3 R_\odot$. Then, only the radius of the companion was fitted.

\begin{figure}
\begin{center}
  \resizebox{0.99\linewidth}{!}{\includegraphics{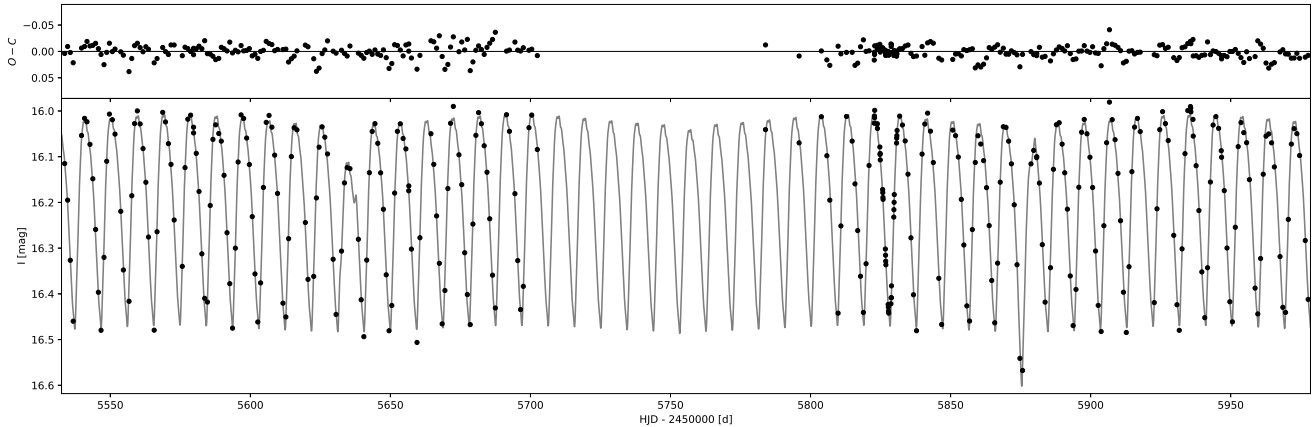}} \\
\caption{OGLE I-band light curve model showing two eclipses of a companion by the Cepheid at HJD 2455635 and HJD 2455875. The residuals are mostly due to a modulation of the pulsational light curve.
\label{fig:8280_model}}
\end{center}
\end{figure}

To obtain an optimal solution and an error estimation a standard Monte Carlo and Markov Chain Monte Carlo (MCMC) sampling was used.  The final light curve model is presented in Fig.~\ref{fig:8280_model}.

To estimate the temperature we have used a semi-empirical transformation from \cite{1993ApJS...86..541C} and the $(V-I)$ color calculated from the corresponding P-L relations \citep{2008AcA....58..293S}, assuming an average combined LMC and Galactic foreground reddening value of $E(B-V)$=0.12 mag \citep{2007ApJ...662..969I}. The transformation is given below:

$$ \log(T_{eff}) = 4.199 - \sqrt{0.04622 + 0.2222(V-I)} $$

We note that the V and I-band measured system magnitudes lie slightly above the respective P-L relations for W Virginis stars, which is consistent with some extra light coming from the companion. Moreover, on the P-L relation for the Wessenheit index, the system also lies above and equally close to the relation, suggesting that the possible intrinsic extinction is not  significantly affecting the system brightness and that the majority of the light indeed comes from the Cepheid.
We thus find it reasonable in this case to use Cepheid magnitudes derived from the P-L relations for the temperature estimation, although it is not yet clear if this approach can be used for all peculiar W Virginis stars.

We derive a temperature of the Cepheid of 5400 K. Using  the system color directly we obtain the slightly higher temperature of 5600 K, which is consistent with our view of the companion being hotter, but significantly fainter, than the Cepheid. The estimated uncertainty is 200 K, which accounts for errors in  the photometry and the parameters of the relation. Some unknown systematical error related to the use of the W Virginis period-luminosity relation for this star may be still present.

The temperature of the companion was  estimated from the residual color after the subtraction of the Cepheid light from the total system brightness and according to the light curve solution. As the system and Cepheid brightness is very similar the uncertainty of the estimated color of the companion is high. The lack of a disk in the light curve model also makes the results regarding the surface brightness ratio quite speculative. The presence of  circumstellar matter may also further affect the solution.

The resulting physical parameters of the OGLE-LMC-T2CEP-211 system components are given in Table~\ref{tab:8280abs}. There are two classes of parameters listed, those mainly based on the orbital solution (which are more robust) and those that rely on the light curve model and photometry (the radii and especially the temperatures) which are prone to strong systematics due to the missing components in the model, and correlations between the parameters.

The masses belong to the first group, and are very well determined, with uncertainties of about $2.3\%$ for the Cepheid ($0.643 \pm 0.015\, M_\odot$) and $1.1\%$ for the companion ($5.67 \pm 0.06\, M_\odot$). We should also not expect problems with the determination of the sum of the radii ($34 \pm 2\, R_\odot$) and indeed it is consistent with what we can simply calculate from the width of the eclipse assuming an edge-on view of the system orbit ($33 \pm 4 R_\odot$). The individual radii and temperatures are highly correlated and are the weakest points of the solution. This is also due to the fact that the companion is probably self-obscured by the disk, which is not taken into account in the model. Using the Cepheid radius obtained with the help of pulsation theory ($25.1\, R_\odot$), yields $R=9 \pm 2\, R_\odot$ for the companion.

If we fit the radius of the Cepheid as a free parameter, we obtain a value of $28.8 \pm 0.7\, R_\odot$. In this solution the companion is smaller ($5.4 \pm 1.0 \, R_\odot$) and the inclination is lower ($86.1^{\circ}$). Because of the presence of the relatively dense circumstellar matter,  in neither case do the radii correspond to the secondary star surface, but rather to the bright part of the whole complex object. As suggested by the H$_\alpha$ profile, what we detect is probably a shell surrounding a smaller early-type star, as typical for Be shell stars. Circumstellar material with a radius of similar size ($\sim$9-10 $R_\odot$) was detected around $\phi$ Persei, a B1.5e shell star \citep{2001A&A...368..471H}.
The radius, temperature and other related quantities (like luminosity) in our solution (Tab.~\ref{tab:8280abs}) are thus probably not directly related to the photosphere of the companion. For example, \citet{2006ApJ...639.1081C} modeled a disk around a Be star with a surface temperature of 19000 K and found that the disk temperature changes from 15000 K to 7000 K depending on the distance from its base.

\begin{deluxetable}{lccc}
\tablecaption{Properties of OGLE-LMC-T2CEP-211. \label{tab:8280abs}}
\tablewidth{0pt}
\tablehead{
\colhead{Parameter} & \colhead{Primary} & \colhead{Secondary (Shell)} & \colhead{Unit}
}
\startdata
spectral type        & G1 II             & ---               &       \\   
mass                 & 0.643 $\pm$ 0.015 & 5.67 $\pm$ 0.06   & $M_\odot$ \\
radius               & 25.1 $\pm$ 0.3    &  (9  $\pm$ 2)     & $R_\odot$ \\
$\log g$             & 1.446 $\pm$ 0.015 & (3.3  $\pm$ 0.2)  & cgs \\ 
temperature          & 5400 $\pm$ (200)  & (7000 $\pm$ 1000) & K\\      
$\log L/L_\odot$     & 2.68 $\pm$ 0.07   & (2.2  $\pm$ 0.3)  & \\
$I_C$ (P-L)          & 16.35 $\pm$ 0.10  & (17.7 $\pm$ 0.4)  & mag \\  
$V$ (P-L)          & 17.30 $\pm$ 0.15  & (18.4 $\pm$ 0.4)  & mag \\  
orbital period       & \multicolumn{2}{c}{242.23$\pm$ 0.06 } & days \\
$T_{pri}$            & \multicolumn{2}{c}{2455754.3 $\pm$ 0.2 }   & days \\
semimajor axis       & \multicolumn{2}{c}{302.6 $\pm$ 1.5}   & $R_\odot$ \\ 
inclination          & \multicolumn{2}{c}{86.5 $\pm$ 1.0}    & degrees \\
$R_1+R_2$            & \multicolumn{2}{c}{34   $\pm$ 2}      & $R_\odot$ \\
\hline
outer ring radius    &                   & 116  $\pm$ 1      & $R_\odot$ \\
inner ring $R_{out}$ &                   &  64 $\pm$ 0.5     & $R_\odot$ \\
inner ring $R_{in}$  &                   &  35 $\pm$ 3       & $R_\odot$ \\
\enddata
\tablecomments{Uncertain observational properties that refer to disk-obscured component (probably a shell star) are given in parentheses.}
\end{deluxetable}

\subsection{Disk modeling} \label{sub:disk}

\begin{figure}
\begin{center}
  \resizebox{0.6\linewidth}{!}{\includegraphics{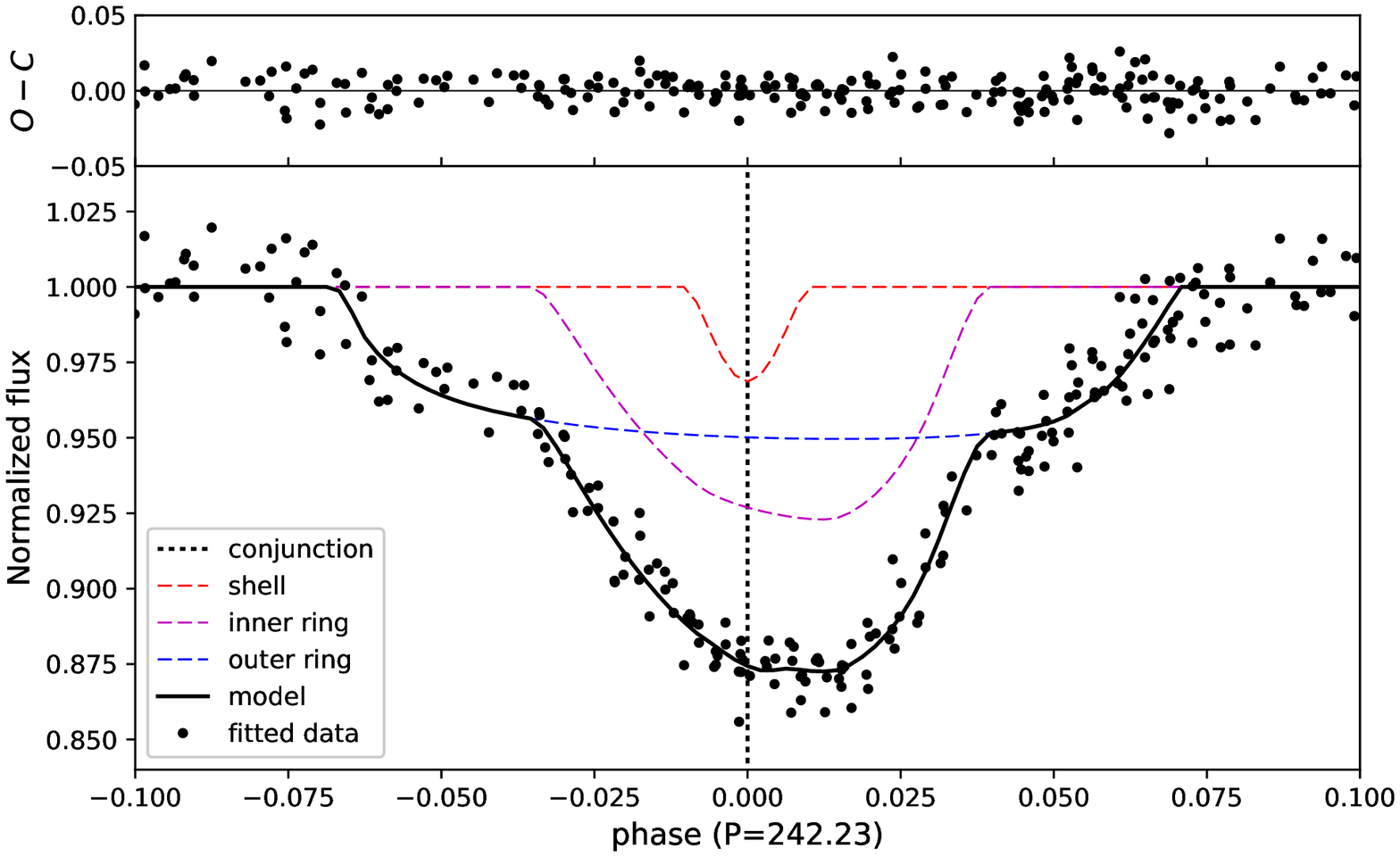}} \\
  \resizebox{0.6\linewidth}{!}{\includegraphics{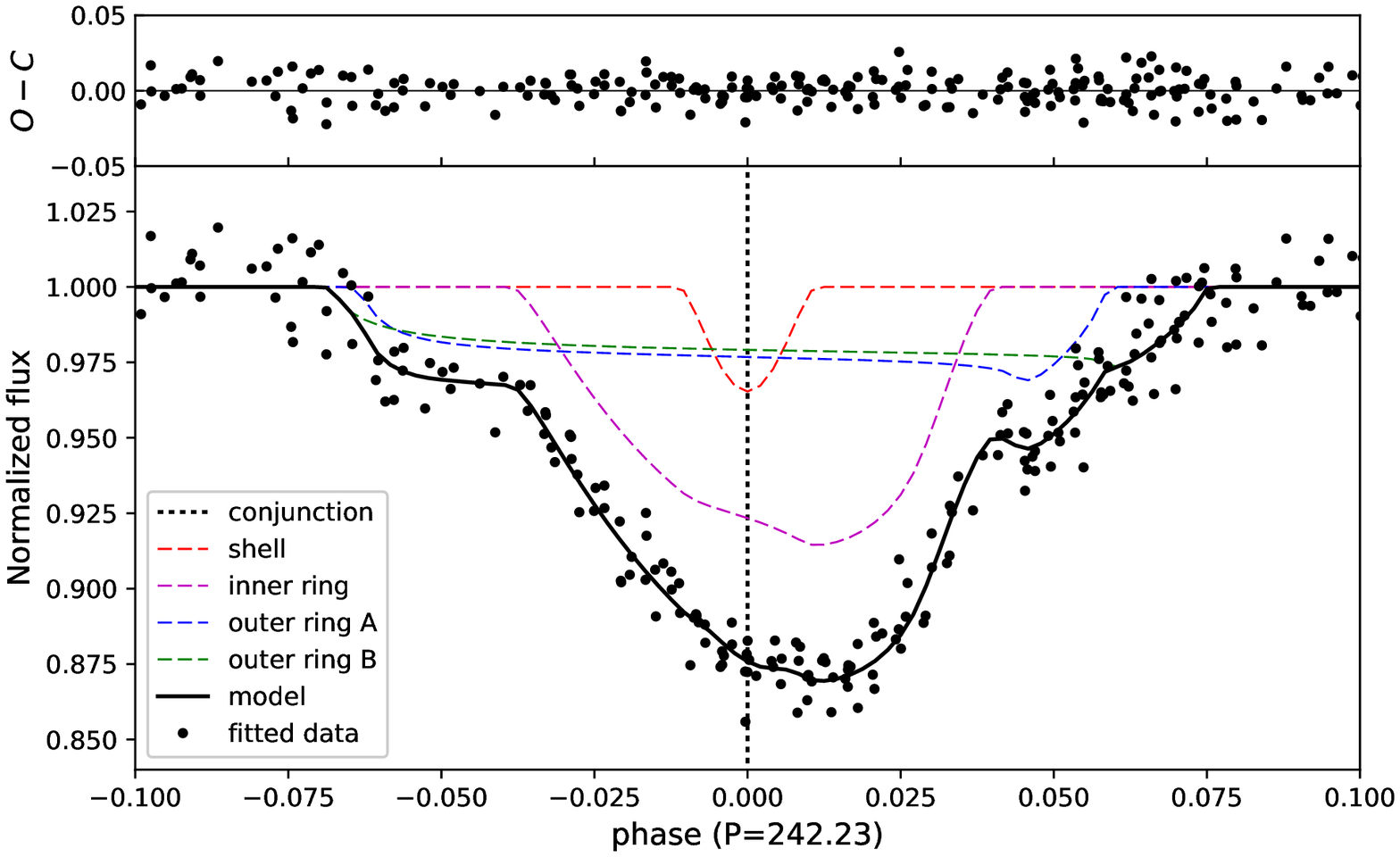}} \\
\caption{{\em top:} A model of the eclipse of the Cepheid by a complex disk consisting of two rings, a broad inner and a narrow outer one. The I-band eclipsing light curve $LC_{ecl}$ is folded with the ephemeris $T_{pri} ($HJD$) = 2455754.27 + 242.23\times $E.  The eclipse minimum is shifted in phase from the moment of conjunction (dotted line). Eclipse shape for each component of the secondary separately is plotted with a dashed line. Note that the shell contribution in the combined model is much smaller because of non-zero projected intersection with the rings. $\vert$ {\em bottom:} The same plot, but with the outer ring divided in two parts. A better fit in this case points to some irregularities in the ring density or a finer structure.
\label{fig:8280_diskecl}}
\end{center}
\end{figure}

We have also performed a separate simple modeling of the disk using the shape of the primary eclipse only. We have tested for a presence of various combinations of a solid disk and up to two rings, taking into account the inner ($R_{in}$) and outer ($R_{out}$) radius of each ring, and their thickness and inclination to the orbital plane. Disk asymmetry was also modeled, represented as a simple translation of the ring center in the orbital plane at any direction. We have assumed the whole disk to be optically thick. For the Cepheid an average radius was used and the limb darkening was not taken into account. The observed data and the fit are shown in the top panel of Fig.~\ref{fig:8280_diskecl}.

\begin{figure}
\begin{center}
  \resizebox{0.99\linewidth}{!}{\includegraphics{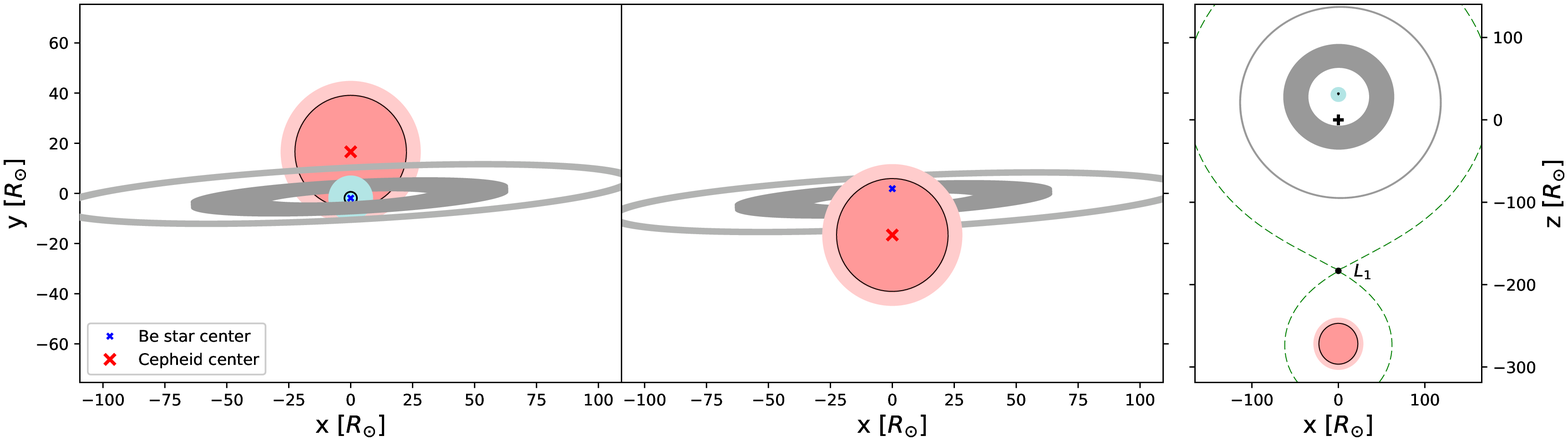}} \\
\caption{
Schematic model of the system. The Cepheid is depicted by the red filled circle, with the minimum radius marked by a black line, and the radius variation range in a lighter color. The companion is depicted by the blue filled circle with a black border, while the shell surrounding it is colored light blue, and the rings of the disk in gray. In the right panel the Roche lobes contours are plotted with a dashed line and the barycenter is marked by $+$.
\label{fig:8280_config}}
\end{center}
\end{figure}

We have obtained the best fit for a model with two rings surrounding the companion. A configuration of the system for the best model is shown in Fig.~\ref{fig:8280_config}.
The inner ring extends from about 35 to 64 $R_\odot$ from the companion center, while the outer ring is located at $\sim{}116\,R_\odot$ and is practically torus-like (with the radial extent comparable to its thickness). The whole disk is slightly tilted from the orbital plane by 1-2$^\circ$.
It is also asymmetric, with a significant extension toward the Lagrangian point $L_1$  and slightly toward the direction of the orbital motion. This feature is most prominent in the case of the outer ring. 

A similar eclipse shape as the one modeled here was observed in the EE Cep system, where a ring structure and an inclined disk model was proposed to explain the observed light curve features \citep{2010ASPC..435..423G}.

The outer ring is possibly formed from the matter lost by the Cepheid due to stellar wind and pulsations and transferred to the orbit around the companion through the $L_1$ point as predicted by \citet{1967AcA....17..297K}, although a dynamical simulation would have to be made to proof this hypothesis. In our simple model, such a stream of mass can be represented as a thin, asymmetric ring which coincides with the stream trajectory in the phase range of the eclipse.
The small waves seen in the residuals between phases 0.04 and 0.08 may indicate some finer features related to the stream that are not included in our two-ring model. We have tried to account for it by splitting the outer ring into two separate ones, adjusting their radii and relative shifts. The best light curve model for such a configuration is presented in the bottom panel of Fig.~\ref{fig:8280_diskecl}. Much smaller residuals indicate that indeed two circular, non-concentric streams of matter located at about 105 $R_\odot$ and 125 $R_\odot$ can explain the eclipse shape. However, as we are probing just a part of the ring that is occulting the Cepheid, we cannot say how the ring structure looks all around the orbit. For this reason we keep as the best one the two-ring model, which gives a slightly worse fit but is more robust. We have to keep in mind, however, that the rings, and especially the outer one, may have a more complex, finer structure.

The origin of the inner ring is not clear. It can be either an accretion or decretion disk, but the gaps between the shell-covered star and the ring, and between the ring and the outer ring, make its existence more enigmatic. In general the shell around Be stars is modeled as a disk \citep{2001A&A...368..471H}. This means that apart from the two rings, probably of accretion origin, there is also an additional decretion disk around the companion with a radius of ~9 $R_\odot$, making the whole structure of the system even more complicated.

Please note that this model accounts only for optically thick parts of the disk, so a solid disk with interchanging thick and thin zones may also explain the light curve variation. Nevertheless, our light curve model is nicely complemented with the information obtained from the analysis of the hydrogen lines. As seen in Fig.~\ref{fig:8280_halpha} and \ref{fig:8280_hbeta}, the maxima of the emission in H$_\alpha$ (at $r = 113$-$122$ R$_\odot$) correspond perfectly to the position of the outer ring, while the maxima of the emission in H$_\beta$ (at $r = 64$ R$_\odot$) correspond to the outer radius of the inner ring -- see Table~\ref{tab:8280abs}.

As seen in Fig.~\ref{fig:8280_config} the system is clearly a detached one, as both stars are well within their corresponding Roche lobes.

\section{Binary evolution analysis} \label{sec:binevol}

The current configuration of the system and especially the masses of components directly point to a binary evolution scenario.
We have performed a binary evolution analysis with the aim to reconstruct the observed parameters of the components and to determine the initial configuration of the system.  As the secondary component is apparently severely obscured by disk and circumstellar matter, which makes impossible a direct determination of its radius and temperature, such analysis may also reveal the expected physical parameters for the hidden companion. Fortunately the masses are precisely determined for both components and serve as very good constraints for the evolutionary models. 

In the course of the evolution of stars in a binary system the validity of the Vogt-Russell theorem (that the structure and evolution of a star is uniquely determined by its mass and chemical composition, \citealt{2006book_modern_aph_and_cosm}) may be altered by the interaction of the components in the system. The Roche-lobe induced mass transfer is one of the most important, affecting the individual components in a way that their structure and evolutionary paths change drastically. Moreover, the combination of the transferred, accreted and lost mass from the components entangles the individual history of the components, so the investigation of the progenitor's properties in post-mass transfer systems always suffers from some amount of degeneracy.

The fate of a binary system after the onset of a mass transfer is a complex combination of an individual structural response of components to the mass change and the angular momentum evolution. In principle, the evolution of the angular momentum (AM) of a system is determined by the initial masses of its components, the orbital period and the mass/AM-loss from the system during the mass transfer.
Although accurate treatment of this process is not yet possible, assuming a simplified AM loss mechanism proposed by \cite{Hurley+2002_binevol} we can track the AM evolution of a system for a given initial mass ratio ($q^i$) and systemic mass loss fraction ($\beta$):

\begin{equation}
\beta=1-\left\vert\frac{\dot{M_a}}{\dot{M_d}}\right\vert\quad 0\leq\beta\leq1,
\label{beta}
\end{equation}

\noindent
where $\dot{M_a}$ is the mass being received by the accretor and $\dot{M_d}$ is the mass being lost by the donor. The mass transfer is conservative if $\beta=0$ and totally non-conservative if $\beta=1$.

Using the simplified AM loss mechanism it is possible to calculate the initial parameters of a system for a given set of $q^i$ and $\beta$ using equations 3 and 4 presented in \cite{Kolbas+2014_binevol}, while to track the structural response of the components, one have to use a stellar evolution code capable of calculating simultaneously and in detail the internal structure of both star during the mass transfer.

To determine the evolutionary status of OGLE-LMC-T2CEP-211, we first calculated sets of initial parameters consistent with the current total mass and AM of the system. The calculations followed the method and conventions presented in \citet{Kolbas+2014_binevol}. The input parameters were: the final masses of the donor and accretor ($M_d^f$, $M_a^f$), the final mass ratio ($q^f = M_a^f/M_d^f$), and the final orbital period ($P^f$). For the initial mass ratio ($q^i = M_a^i/M_d^i$) and the systemic mass loss fraction $\beta$ assumed for the given model we calculated the initial masses of the donor and accretor ($M_d^i$, $M_a^i$), and the initial orbital period ($P^i$). We have repeated these calculations for 134 sets of values of $q^i$ and $\beta$. Finally, we created binary evolution grids starting from all the initial configurations for stars with an LMC metallicity, Z=0.008.
We carried out the calculations using the Cambridge version of {\tt STARS} evolution code \citep{Eggleton1971,Eggleton1972, Pols1995_stellar_modelling, Stancliffe2009}. More details on building this kind of binary evolution grids and performed calculations can be found in \citet{2018_Dervisoglu}.
The results of these calculations, i.e. the evolutionary tracks, were then compared with the current configuration of the system. 

We searched for the best fitting binary evolution model by means of  $\chi^2$ minimization comparing the results with the observed, absolute properties of the both components of the system. However, since the accretor is highly obscured by the disk, we excluded its radius and temperature from our analysis. We define the $\chi^2$ function as:
\begin{equation}
\chi^2 = \Sigma \frac{(X_{obs} - X_{mod})^{2}}{\sigma_{obs}^{2}},
\end{equation}
where $X_{obs}$ and $X_{mod}$ are, respectively, the observed and modeled: donor mass $M_d$, radius $R_d$, temperature $T_d$ and accretor mass $M_a$. The orbital period is not included here because during the mass transfer it is a function of masses of the components and $\beta$, i.e. $P \cdot M_t^2 \cdot M_a^3 \cdot M_d^{1-\beta}= const$.

\begin{figure}
\begin{center}
  \resizebox{0.6\linewidth}{!}{\includegraphics{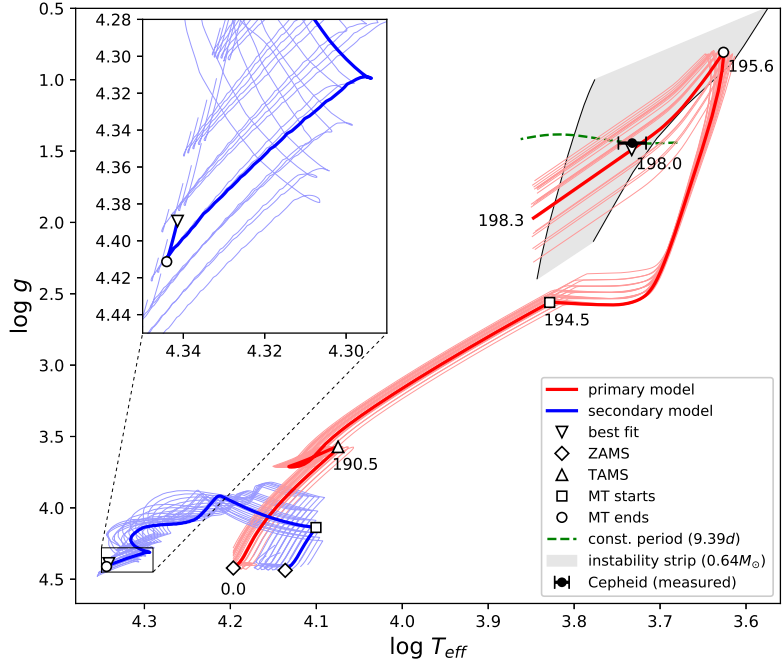}}
\caption{ Evolutionary tracks for both components and the instability strip (IS) for stars of $M=0.64$ $M_{\odot}$ with   the recent history of the secondary shown in the inset. Ages in million years (My) are shown for important evolutionary stages. The mass transfer has  recently finished and the Cepheid is now moving towards lower temperatures passing through the IS. This phase is relatively short, lasting from 0.3 to 3 My depending on the model. Only the total mass and current parameters of the Cepheid component were used in modeling as the companion is heavily obscured by the disk.
\label{fig:8280_binevol}}
\end{center}
\end{figure}

The best representative model was found for $q^{i} = 0.80$ and  $\beta = 0.0$ with the corresponding initial parameters of the system being: $M_d^i = 3.52 M_\odot$, $M_a^i = 2.82 M_\odot$, and $P^{i} = 12.0$ days. For this model the minimum $\chi^2=9.85$ was found for $M_d=0.715$, $T_d=5410$ , $L_d=484$, and $R_d=25.12$ for the mass donor and $M_a=5.625$, $T_a=21950$ , $L_a=1310$ and $R_a=2.506$ for the mass accretor. The systemic parameters of this best-fit model are $q=7.87$, $P_{orb}=180.8$, and the age of the system is $t=198.0$ million years. We also calculated the confidence limit of best fitting model by propagating the observed errors into $\chi^2$ function, obtaining the value $\chi^2_{lim} = 32.4$. From all calculated models 24 fall within this confidence limit, i.e. have $\chi^2 \leq \chi^2_{lim}$ . All these valid models are presented in the electronic version of the manuscript and plotted on the $\log g$ - $\log T_{eff}$ (Kiel) diagram with the observed position of the components in Fig.~\ref{fig:8280_binevol}.  A subset of them is shown in Table~\ref{tab:8280_modevol_sum}, together with the best and the mean model. Standard deviation of the parameter values for valid models is also given, and is used as an estimation of the uncertainties of the best model.

Nonetheless, caution is required in interpreting the presented results and confidence intervals, as binary evolution models are not yet very exact, missing much complex physics regarding the mass transfer and angular momentum evolution. They are still in active development as our knowledge of these processes increases. But even with the systematic errors that may be present in these models, they are still our best way to study the history of the observed binaries, providing us with plausible parameter sets and indicating the current and past important stages of the evolution.

\begin{deluxetable}{lc ccc ccc ccc ccc ccc}
\tabletypesize{\scriptsize}
\tablecaption{Evolutionary models. \label{tab:8280_modevol_sum}}
\tablewidth{0pt}
\tablehead{ \colhead{model}
 & \colhead{$q^i$} & \colhead{$\beta$}  & \colhead{$M_d^i$} & \colhead{$M_a^i$} & \colhead{$M_{tot}$} & \colhead{$P^i$} & \colhead{$q^f$}
 & \colhead{$M_d^f$} & \colhead{$\log T_d$} & \colhead{$\log L_d$} & \colhead{$\log R_d$}
 & \colhead{$M_a^f$} & \colhead{$\log T_a$} & \colhead{$\log L_a$} & \colhead{$\log R_a$}
 & \colhead{$P^f$} 
}
\startdata
\hline
 1    & 0.909 & 0.000 & 3.321 & 3.019 & 6.340 & 11.664& 7.716 & 0.727 & 3.678 & 2.461 & 1.398 & 5.613 & 4.324 & 3.173 & 0.463 & 172.759 \\
 2    & 0.909 & 0.050 & 3.393 & 3.085 & 6.478 & 12.614& 7.778 & 0.723 & 3.690 & 2.510 & 1.399 & 5.621 & 4.325 & 3.177 & 0.462 & 178.260 \\
 3    & 0.909 & 0.100 & 3.469 & 3.154 & 6.623 & 13.657& 7.818 & 0.720 & 3.707 & 2.578 & 1.399 & 5.628 & 4.326 & 3.179 & 0.461 & 182.606 \\
\dots & \dots & \dots & \dots & \dots & \dots & \dots & \dots & \dots & \dots & \dots & \dots & \dots & \dots & \dots & \dots & \dots   \\
\hline
best  & 0.80  & 0.0   & 3.52  & 2.82  & 6.34  & 12.0  & 7.87  & 0.715 & 3.73  & 2.68  & 1.400 & 5.625 & 4.341 & 3.12  & 0.40  & 180.8 \\
mean  &	0.83  & 0.08  & 3.60  & 2.99  & 6.59  & 13.6  & 7.80  & 0.722 & 3.75  & 2.76  & 1.401 & 5.625 & 4.340 & 3.14  & 0.41  & 181.2 \\
stdev &	0.06  & 0.07  & 0.12  & 0.19  & 0.22  &  1.5  & 0.05  & 0.004 & 0.03  & 0.14  & 0.002 & 0.005 & 0.009 & 0.03  & 0.03  &   3.2 \\
\enddata
\tablecomments{Extended version of this table with all the considered binary evolution models, plotted in Fig.~\ref{fig:8280_binevol}, is available in the online version of the manuscript. As a summary, values for the best and mean model with the standard deviation of the parameters are given here. The meaning of the indices is: i -- initial, f -- final, d -- donor, a -- accretor.}
\end{deluxetable}

\begin{deluxetable}{l|cc|c|c}
\tablecaption{Results from binary evolution analysis. \label{tab:8280_binevol}}
\tablewidth{0pt}
\tablehead{
\colhead{Parameter} & \colhead{Initial}  & \colhead{Current}  & \colhead{Measured} & \colhead{Unit}
}
\startdata
\hline
                       \multicolumn{5}{c}{System}                      \\
mass ratio           &   0.8 $\pm$ 0.06  &  7.87 $\pm$ 0.05  & 8.82 $\pm$ 0.14   & \\
orbital period       &  12.0 $\pm$ 1.5   & 180.8 $\pm$ 3.2   & 242.23 $\pm$ 0.06 & days \\
$\beta$              &                   &   0.0 $\pm$ 0.07  &                   & \\
\hline
                       \multicolumn{5}{c}{Primary component / currently Cepheid}                      \\
spectral type        &        B5V        &       G1 II       &       G1 II       & \\
$M_1$                &  3.52 $\pm$ 0.12  & 0.715 $\pm$ 0.004 & 0.643 $\pm$ 0.015 & $M_\odot$ \\
$R_1$                &  1.91 $\pm$ 0.13  & 25.12 $\pm$ 0.12  & 25.1  $\pm$ 0.3   & $R_\odot$ \\
$\log g$             &  4.42 $\pm$ 0.06  & 1.491 $\pm$ 0.005 & 1.446 $\pm$ 0.015 & cgs \\
$T_1$                & 15700 $\pm$ 450   &  5410 $\pm$ 370   &  5400 $\pm$ 200   & K \\
$\log L_1/L_\odot$   &  2.30 $\pm$ 0.05  &  2.65 $\pm$ 0.06  &  2.68 $\pm$ 0.07  & \\
\hline
                       \multicolumn{5}{c}{Secondary component}                                        \\
spectral type        &        B7V        &       B2 V        &      ---          & \\
$M_2$                &  2.82 $\pm$ 0.19  & 5.625 $\pm$ 0.005 & 5.67 $\pm$ 0.06   & $M_\odot$ \\
$R_2$                &  1.68 $\pm$ 0.12  &  2.51 $\pm$ 0.17  &   (9 $\pm$ 2)     & $R_\odot$ \\
$\log g$             &  4.44 $\pm$ 0.07  &  4.39 $\pm$ 0.06  &      ---          & cgs \\
$T_2$                & 13700 $\pm$ 530   & 22000 $\pm$ 450   & (7000 $\pm$ 1000) & K \\
$\log L_2/L_\odot$   &  1.95 $\pm$ 0.10  & 3.117 $\pm$ 0.011 &      ---          & \\
\enddata
\tablecomments{Current and initial states of the system are given for the evolutionary model. $R_2$ and $T_2$ refer to the star in case of the binary evolution model and to the central, bright region (probably a shell) of the star+disk component in case of the measurement.}
\end{deluxetable}

In our best fitting model, the mass transfer from the current Cepheid component to its companion starts 194.5 million years after ZAMS (Zero Age Main Sequence) and it takes 1 million years for the transfer to complete. The current age of the components is about 198 My and the mass transfer has finished about 2.5 million years ago.
The primary component has mass $M_1 = 0.715 M_\odot$ and a highly evolved helium core. The total mass of helium is 0.66 $M_\odot$, while the hydrogen is present only in the envelope. The expected surface mass fraction of hydrogen and helium are $X=0.42$ and $Y=0.57$, respectively.

The mass of the primary component in this model (see Model 1 in Table~\ref{tab:8280_binevol}) is higher than the measured one, but the mass loss from winds was intentionally disabled to track the mass transfer evolution only. The mass excess $\Delta M = 0.072 M_\odot$ can be removed from the primary with mass loss rates of the order of $10^{-7} - 10^{-8}$ $M_\odot$/year during the time after the mass transfer has finished. Such mass loss rates are common for highly evolved stars with low surface gravity \citep{1986ApJ...311..731K}, and the pulsations may further enhance the process \citep{2008ApJ...684..569N}.

To compare the model with observations for the current Cepheid we have enabled mass loss due to wind $\dot{M}_{wind} = 2.7 \times 10^{-8} M_{\odot}/yr$ after the mass transfer phase. In this  model the mass of the primary ($M_1 = 0.66 M_\odot$) and the orbital period ($P_{orb} = 251.3)$ are consistent with the observations within the simplified modeling uncertainty limits.

In Fig.~\ref{fig:8280_binevol} the evolutionary tracks for both components are presented together with the observational data. The most important phases during the binary evolution, i.e. ZAMS, end of the main sequence evolution, and mass transfer beginning and end are also marked. 

The binary evolution model strongly suggests that the currently more massive  companion is much smaller and hotter (B2 type) than the bright component detected during the light curve modeling. This further strengthens our conclusion of the presence of a larger ($\sim 9$ $R_\odot$) shell surrounding the fast-rotating and much smaller central B-type star, ie. a typical Be shell star. The star itself is probably highly distorted by the rapid rotation. Such a configuration explains all of the observations and is also consistent with the current view on  Be shell stars.

\section{Pulsational properties} \label{sec:pulsprop}

As mentioned in the previous section, the Cepheid  is almost totally stripped of hydrogen at the current epoch. The star has a helium core and the small fraction of hydrogen can be found only in the envelope, with about $42\%$ of hydrogen at the surface level. Such an unusual composition could influence the pulsational properties of the Cepheid, which we verify with pulsation calculations. We use the linear and non-linear pulsation codes of \citet{sm08}. These are lagrangian, one-dimensional codes which implement the turbulent convection model of \citet{kuhfuss}. In the calculations we use OPAL opacity tables \citep{opal}, supplemented at  low temperatures with the \citet{af05} opacity data. For the distribution of heavier elements we adopt a scaled solar mixture as given by \citet{a09}. To study the effects of the strongly decreased hydrogen content (and consequently strongly increased helium content), we consider two chemical compositions of the models: with $X=0.42$ and $Z=0.01$ (model X42), as inferred from evolutionary modeling, and with $X=0.71$ and $Z=0.01$ (model X71). For convective parameters we adopt the values  previously used in the modeling of type II Cepheids by \citet{sm12} and \citet{s16}. The exact values can be found in Tab.~2 of \citet{sm12} (set P1). Physical parameters are fixed to the values from Tab.~\ref{tab:8280abs}

At the linear level, for both models  X42 and X71  we obtain satisfactory results. The period of the fundamental mode is barely sensitive to the hydrogen content, the difference for the two adopted compositions is $\approx 0.03$\,d. For both compositions the fundamental mode is linearly unstable, i.e. the Cepheid is located within the instability strip. The boundaries of the instability strip  for model X42 are plotted in Fig.~\ref{fig:8280_binevol}. We note that for model X42, the convective zones  are more extended and carry less radiative flux than for model X71. As a result the driving mechanism is less efficient and the absolute value of the growth rate is smaller for model X42.

The evolutionary track of the initial primary is located inside the  boundaries of the instability strip ever since it finished transferring mass to the initial secondary. This means that the current Cepheid could start pulsating quite soon after the mass transfer phase was finished. This may be an important factor responsible for enhancing the mass loss \citep{2008ApJ...684..569N}, which is necessary to explain the current mass of the Cepheid.

\begin{figure}
\begin{center}
  \resizebox{0.49\linewidth}{!}{\includegraphics{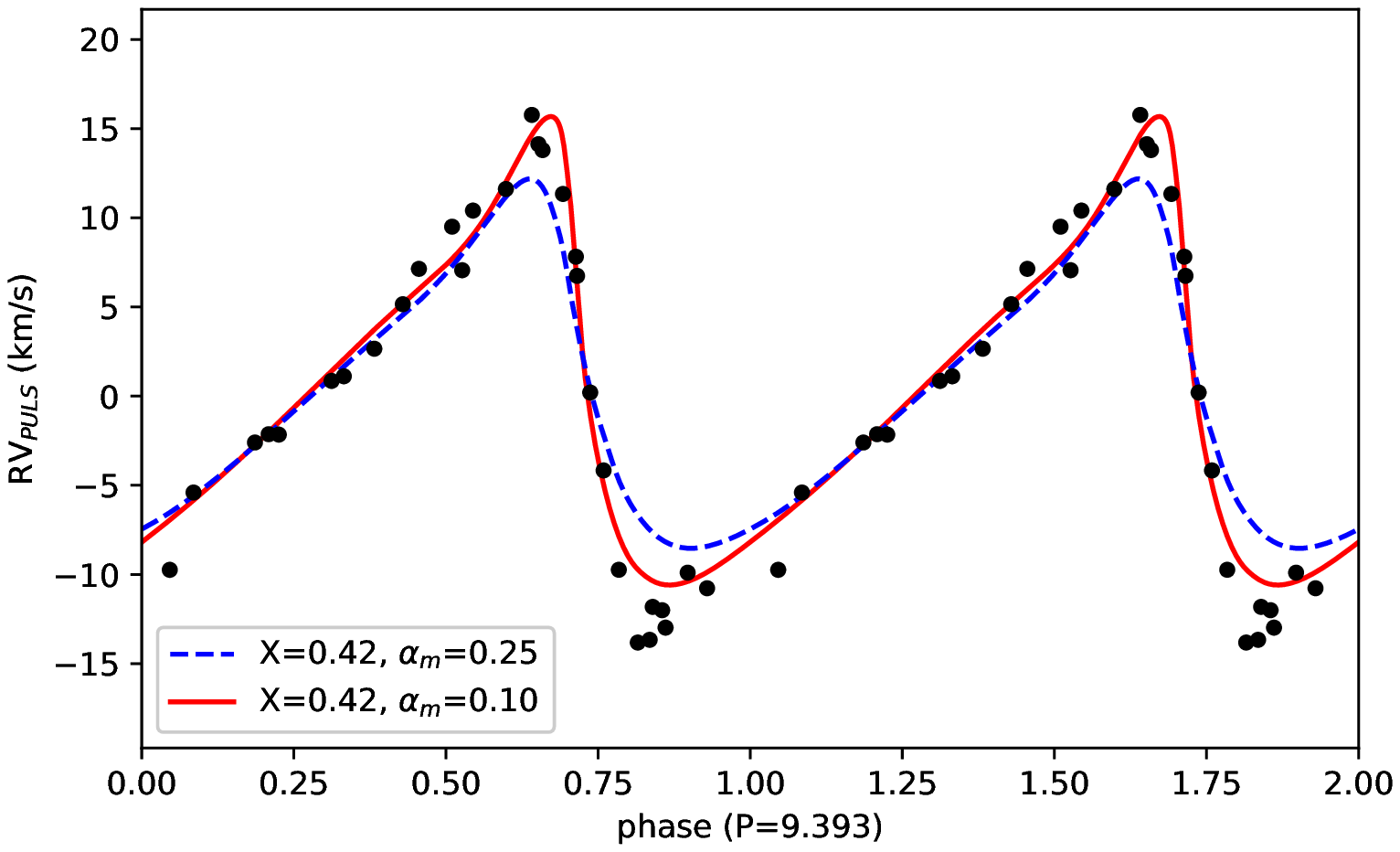}}
  \resizebox{0.49\linewidth}{!}{\includegraphics{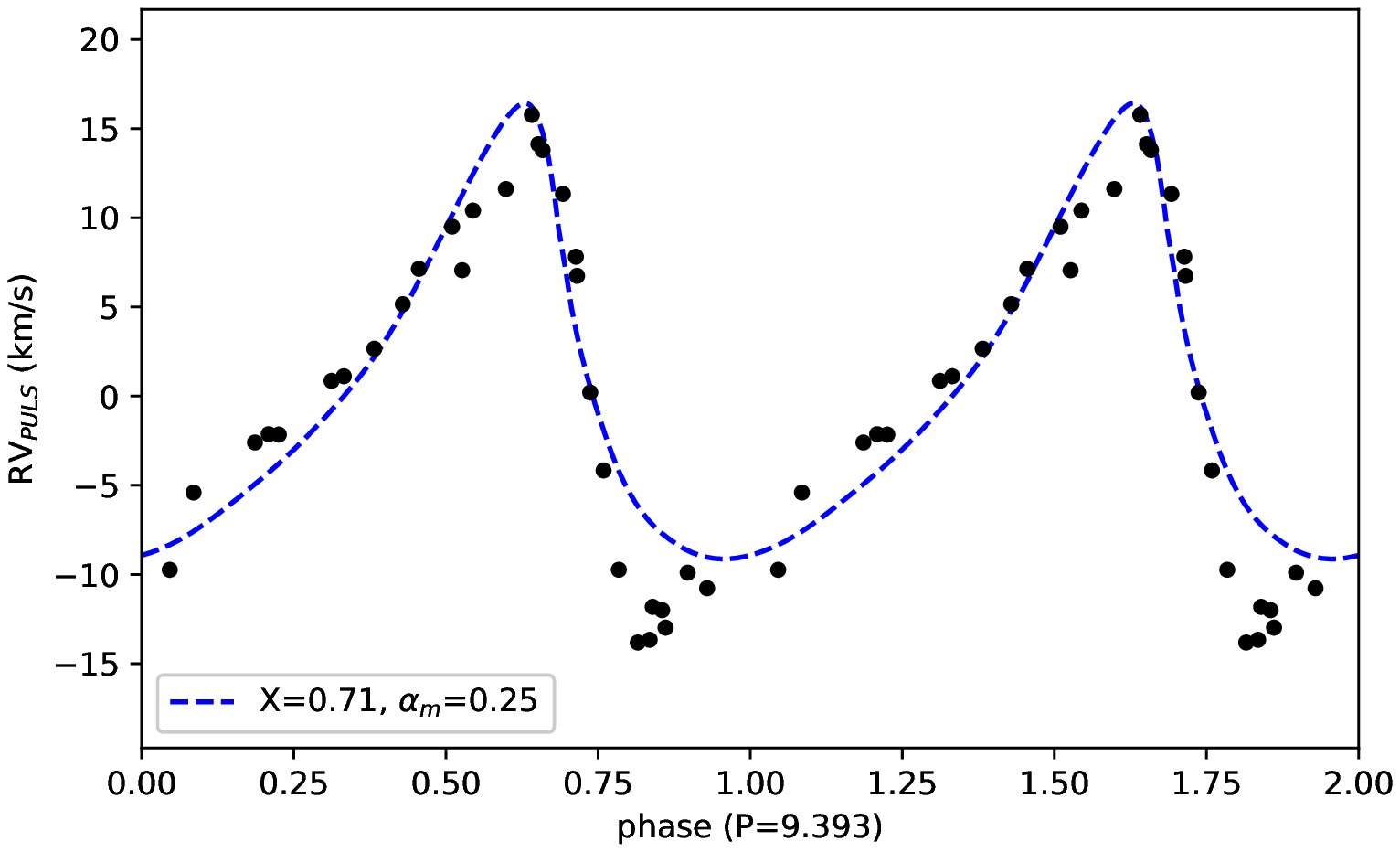}}
\caption{ Comparison of observed and computed radial velocity curves. The values of p-factor 1.25 and 1.3 were used for model X42 and X71, respectively. For all model RV curves, a shift in phase was applied to obtain the best possible fit to the data.
\label{fig:8280_rvcomp}}
\end{center}
\end{figure}

In Fig.~\ref{fig:8280_rvcomp} we show the pulsation radial velocity curves computed with the non-linear code for the two chemical compositions: $X=0.42$ in the left panel and $X=0.71$ in the right panel. To allow comparison with the observed radial velocities the calculated pulsational velocities were divided by the projection factor and shifted in phase to obtain the best fit. The p-factor values in a narrow range of 1.25-1.35 were considered. We obtained $p=1.25$ for model X42 and $p=1.3$ for model X71. For model X42 the amplitude is smaller than the observations. To increase the amplitude we have decreased the eddy-viscous parameter from $\alpha_m=0.25$ to $\alpha_{m}=0.10$ (models X42$_{0.25}$ and X42$_{0.10}$). The resulting radial velocity curve is plotted with a dashed line and satisfactorily reproduces the observed radial velocity curve. The only significant difference is that the maximum expansion velocity is marginally too small. For model X71 the amplitude roughly matches the observations, but the overall agreement with the observed radial velocity curve is poor, and the change of $\alpha_m$ does not help. The corresponding Fourier decomposition coefficients for the observed and modeled RV curves are presented in Table~\ref{tab:8280_fourier_rv}. Note that these coefficients do not depend on the used p-factor and the phase shift. It is clear that with $X=0.42$ the models are a better fit to the observed radial velocity curve, we note however that more detailed modeling, in particular a consideration of other sets of convective parameters, is needed for more definite conclusions.

Unfortunately, with the available pulsation codes we cannot reliably model the light variation in luminous, low-mass type II Cepheids that are strongly non-adiabatic pulsators. This is due to the simplified treatment of the radiation transfer, for which a simple diffusion approximation is used in the codes. While dynamical properties such as radius and radial velocity variation are well captured, the light variation is not -- for longer periods, the amplitude is too large and spurious spikes are present in the light curves. We refer the reader to \cite{s16} for further details. Here, we  note that for both models X42 and X71 the ascending branch of the light curve is steeper, as is observed in peculiar W Vir stars.

\begin{deluxetable}{lcc cccc}
\tablecaption{Fourier coefficients for RV curves. \label{tab:8280_fourier_rv}}
\tablewidth{0pt}
\tablehead{
\colhead{model} & \colhead{X} & \colhead{$\alpha_{m}$} & \colhead{$\phi_{21}$} & \colhead{$\phi_{31}$} & \colhead{$R_{21}$} & \colhead{$R_{31}$}
}
\startdata
\hline
data         & 0.42 & ---  & 4.83   & 3.60 &   0.454 & 0.185 \\
X42$_{0.25}$ & 0.42 & 0.25 & 4.98   & 3.89 &   0.380 & 0.190 \\
X42$_{0.10}$ & 0.42 & 0.10 & 4.88   & 3.60 &   0.450 & 0.273 \\
X71$_{0.25}$ & 0.71 & 0.25 & 5.48   & 4.78 &   0.370 & 0.165 \\
\enddata
\end{deluxetable}

\pagebreak

\section{Summary and discussion} \label{sec:summary}

Our analysis of brightness, period and the light curve shape confirmed the classification of the pulsating star in the OGLE-LMC-T2CEP-211 system as a peculiar W Virginis, type II Cepheid. This classification is further enforced by the measured large relative radius change,  almost $25\%$ of the minimum radius, and its position on the HR diagram. It was also shown that the star is a genuine member of an eclipsing binary system with the orbital period of 242.2 days and a very high mass ratio ($q=8.82$).

We have presented the first direct and dynamical measurement of the mass of a star of this type. The measured mass of $0.64 \pm 0.02 M_\odot$ is close to the masses expected for Type II Cepheids ($0.5-0.6 M_\odot$) according to pulsational and evolutionary theory studies \citep{2002PASP..114..689W,1997A&A...317..171B,2016CoKon.105..149B}. This indicates that peculiar W Virginis stars are  low-mass pulsators similar to W Virginis stars although  with a different evolutionary history.

Due to the limitations imposed by the data, it was not possible to directly determine the radius of the Cepheid from the modeling, but it could be well constrained  with the help of pulsation theory. We have obtained a value of $25.1 \pm 0.3 R_\odot$, for which a star of this mass  pulsates with the observed period in the fundamental mode. 

The companion to the Cepheid is significantly if not completely obscured by an optically thick circumstellar disk in an analogue with the W Crucis \citep{2006A&A...454..855P} or the $\beta$ Lyrae \citep{2013MNRAS.432..799M} systems, and its physical parameters could not be directly determined, with the exception of the mass ($5.67 \pm 0.06 M_\odot$). Nevertheless, it is certain that the companion is significantly hotter and smaller than the Cepheid.

Although the pulsating components are in many ways different, the OGLE-LMC-T2CEP-211 system is very similar to OGLE-LMC-T2CEP-098 \citep{t2cep098apj2017} considering the current high mass ratio and the implausible evolutionary states assuming  single star evolution for both components separately. The simplest solution for both systems is the mass reversal scenario due to mass transfer as in the solution of the Algol Paradox, which would explain the existence of the system in the current configuration.

We showed that a system initially composed of 3.5 and 2.8 solar-mass stars and a period of about 12 days  would produce a hydrogen-depleted low-mass star (similar to the observed Cepheid) and a hot main-sequence star ($R=2.5$ $R_\odot$) of approximately spectral type B2 after about 200 million years of binary evolution and a significant mass transfer. Although the comparison is limited by the presence of a circumstellar disk, this configuration is consistent with our measurements,  especially of the masses of the components. However a modest and reasonable mass loss due to wind from the current Cepheid is necessary to explain the current orbital period, which is a sensitive function of the mass ratio.

In the light curve analysis we detected a bright part of the companion star+disk component, but it was larger ($R=9$ $R_\odot$) than the size of the B-type dwarf companion expected from the binary evolution model. Together with other indications, including the shape of the H$_\alpha$ line, we conclude that the companion is a Be shell star, with a $\sim$9 $R_\odot$ decretion disk.

Our modeling of the primary eclipse resulted in a detection of an extended disk composed of two (or more) rings around the companion, with the inner extending from about 35 to 64 $R_\odot$ and the outer being located at a radius of about 116 $R_\odot$. The radii of these rings coincide with the emission maxima in H$_\alpha$ (outer ring) and H$_\beta$ (inner ring).

The presence of the disk around the companion is interesting as the system is not semi-detached, and neither star fills its corresponding Roche lobe. Its presence can be explained in three non-exclusive ways. It could be formed from the mass transferred from the current Cepheid in a past stage of the evolution or from the mass being lost now by the Cepheid due to stellar wind and pulsations, with the latter being the most probable scenario for the outer ring. A decretion disk can also be formed from the mass ejected from a fast-rotating companion as typical for Be stars \citep{1991MNRAS.250..432L,2013A&ARv..21...69R}. A small disk of this origin was detected around the companion, but it is unclear if the inner ring could be formed in the same manner. The origin of the inner ring is thus the most enigmatic, with a possibility that it is being fed from both sides.
It is also uncertain whether or not the current mass loss rate of the Cepheid is large enough to feed the disk with sufficient matter, although it is possibly the only source of mass to populate the outer ring, which presence is consistent with the prediction of \citet{1967AcA....17..297K} for mass transfer through the $L_1$ point. This complex configuration informs the discussion of the possible disk feeding mechanisms \citep{2000ASPC..214..617P,2013A&ARv..21...69R} in similar binary systems.

According to \citet{1977ApJ...216..822P} the radius of the maximum possible streamline in an accretion disk in a system with the mass ratio $q=9$ is 137 $R_\odot$ and the radius of the maximum stable streamline is about 130 $R_\odot$ along the line connecting the components. In our model the size of the disc is close to, but below both of these limits, even for the three-ring model for which $R_{outer} = 125$ $R_\odot$.

The pulsational radial velocity curve obtained from a pulsation theory model for a hydrogen-deficient star (X=0.42) with the observed parameters agrees very well with the observed radial velocity curve. The pulsation velocities are not well fit for models with a typical hydrogen abundance (X=0.71). This confirms the results obtained from the binary evolution analysis that the  pulsating star is a low-mass product of mass transfer composed mostly of helium, bound in the binary system with an early-type companion.
\citet{2008AcA....58..293S} found that peculiar W Vir stars are in general bluer and lie above the P-L relation for type II Cepheids, consistent with this picture. This supports the conclusion that the majority, if not all, of pWVir stars are products of a similar binary evolution.
The difference in color and brightness would then depend on the inclination of the orbit (disks seen face-on would not occult the central star) and also on the luminosity and temperature of the early-type companion.

We found OGLE-LMC-T2CEP-211 to be similar to the AU Mon system \citep{2014PASP..126..821M}, with the exception that for  OGLE-LMC-T2CEP-211 the initial period was longer and the mass transfer has  finished, leaving a less massive donor on a much wider orbit around a very massive companion. The future of the AU Mon system may be similar (i.e. a low-mass, hydrogen-deficient star and a massive early-type star), although a separate study would be necessary to determine if the low-mass star would pulsate and as a what kind of a variable. On the other hand, the more distant future of both systems may be a typical Be star with the secondary too faint to be easily detectable as suggested by \citet{2012ASPC..464..213M} and discussed by \citet{2018ApJ...853..156W}. Our results also support the conclusion of \citet{2013ApJ...764..166D} that it is possible that all early-type Be stars result from binary interactions.

The intention of this work was to present the precise and accurate mass determination for the Cepheid and its companion as well as the evolutionary state of this typical example of a peculiar W Virginis star.  The complicated nature of the system, including the presence of a complex  disk, and the limited photometric data gathered so far, meant that a direct solution and precise absolute physical parameters for both components of the system could not be derived. More and better quality photometry together with more complex modeling are both necessary for a better understanding and characterization of this very interesting and challenging system which may hold the clues for a better understanding of the whole class of peculiar W Virginis stars.

\acknowledgments

We gratefully acknowledge financial support for this work from the Polish National Science Center grant SONATA 2014/15/D/ST9/02248. WG and GP gratefully acknowledge support from the Chilean Centro de Astrofísica y Tecnológicas Afines (CATA) BASAL grant AFB-170002. WG also acknowledges support from the Chilean Ministry of Economy, Development and Tourism's Millennium Science Initiative through grant IC120009 awarded to the Millennium Institute of Astrophysics MAS. AD is financially supported by the Croatian Science Foundation through grant IP 2014-09-8656. The OGLE project has received funding from the Polish National Science Centre grant MAESTRO no. 2014/14/A/ST9/00121. The research leading to these results has received funding from the European Research Council (ERC) under the European Union's Horizon 2020 research and innovation program (grant agreement No 695099).

This work is based on observations collected at the European Organisation for Astronomical Research in the Southern Hemisphere under ESO programmes: 096.D-0425(A), 098.D-0263(A) and 100.D-0399(A). We thank ESO, Carnegie, and the CNTAC for generous allocation of observing time for this project. We would also like to thank the support staff at the ESO Paranal observatory and at the Las Campanas Observatory for their help in obtaining the observations, as well as the observers from the OGLE project for dedication of their time for collecting the photometric data. BP would also like to thank Kre{\v s}imir Pavlovski for useful discussions regarding the studied system.

This research has made use of NASA's Astrophysics Data System Service.

\vspace{5mm}
\facilities{VLT:Kueyen (UVES), Magellan:Clay (MIKE)}

\software{
\texttt{RaveSpan} \citep[][\url{https://users.camk.edu.pl/pilecki/ravespan/index.php}]{t2cep098apj2017}, \\
\texttt{JKTEBOP} \citep[][\url{http://www.astro.keele.ac.uk/~jkt/codes/jktebop.html}]{jktebop2004southworth},\\
\texttt{ESO Reflex} \citep[][\url{http://www.eso.org/sci/software/esoreflex/}]{2013A&A...559A..96F}, \\
\texttt{STARS} \citep[][\url{http://www.ast.cam.ac.uk/~stars}]{Stancliffe2009}
}

\bibliographystyle{aasjournal}
\bibliography{cep8280paper}

\listofchanges

\end{document}